\newif\ifarxiv
\arxivtrue

\ifarxiv
  \documentclass[a4paper,11pt]{article}
  \usepackage[scale={0.8,0.9},centering,includeheadfoot]{geometry}
\else
  \documentclass{ar-1col}
\fi
\usepackage{graphicx}
\usepackage{refcount}
\usepackage[colon,longnamesfirst]{natbib}
\usepackage{subfig}
\usepackage{verbatim}

\newcommand{\url}[1]{\texttt{#1}}

\ifarxiv
\else
  \jname{Xxxx. Xxx. Xxx. Xxx.}
  \jvol{AA}
  \jyear{YYYY}
  \doi{10.1146/((please add article doi))}
\fi

\usepackage{xspace,amstext,amsmath}
\def\Fig#1{\textbf{Figure~\ref{#1}}}

\def\Sec#1{\textbf{Section~\ref{#1}}}

\def\eref#1{(\ref{#1})}

\def\mm#1{\ensuremath{\boldsymbol{#1}}} 

\def\RINLA{\texttt{R-INLA}\xspace}

\ifarxiv
  \def\affil#1{\let\thefootnote\relax\footnote{#1}}
  \newenvironment{keywords}{\begin{quote} \textbf{Keywords:}}{\end{quote}}
  \let\cite\citet
\fi

\begin{document}

\bibliographystyle{apalike}

\ifarxiv\else
\markboth{Rue et al.}{Bayesian computing with INLA} \fi

\title{Bayesian Computing with INLA: A Review}

\author{H{\aa}vard Rue$^1$, Andrea Riebler$^1$, Sigrunn H.\
    S{\o}rbye$^2$,\ifarxiv\\ \fi Janine B.\ Illian$^3$, Daniel P.\
    Simpson$^4$ and Finn K.\ Lindgren$^5$
    \affil{$^1$ Department of Mathematical Sciences, Norwegian
        University of Science and Technology, N-7491 Trondheim,
        Norway; email: hrue@math.ntnu.no}
    \affil{$^2$ Department of Mathematics and Statistics, UiT The
        Arctic University of Norway, 9037 Troms{\o}, Norway}
    \affil{$^3$ Centre for Research into Ecological and Environmental
        Modelling, School of Mathematics and Statistics, University of
        St Andrews, St Andrews, Fife KY16 9LZ, United Kingdom}
    \affil{$^4$ Department of Mathematical Sciences, University of
        Bath, Claverton Down, Bath, BA2 7AY, United Kingdom}
    \affil{$^5$ School of Mathematics, The
        University of Edinburgh, James Clerk Maxwell Building,
        The King's Buildings, Peter Guthrie Tait Road, Edinburgh, EH9 3FD,
        United Kingdom}
}

\ifarxiv \date{\today} \maketitle\fi

\begin{abstract}

    The key operation in Bayesian inference, is to compute
    high-dimensional integrals. An old approximate technique is the
    Laplace method or approximation, which dates back to Pierre-Simon
    Laplace (1774). This simple idea approximates the integrand with a
    second order Taylor expansion around the mode and computes the
    integral analytically. By developing a nested version of this
    classical idea, combined with modern numerical techniques for
    sparse matrices, we obtain the approach of \emph{Integrated Nested
        Laplace Approximations} (INLA) to do approximate Bayesian
    inference for \emph{latent Gaussian models} (LGMs). LGMs represent
    an important model-abstraction for Bayesian inference and include
    a large proportion of the statistical models used today. In this
    review, we will discuss the reasons for the success of the
    INLA-approach, the \RINLA package, why it is so accurate, why the
    approximations are very quick to compute and why LGMs make such a
    useful concept for Bayesian computing.

\end{abstract}


\ifarxiv\else\maketitle\tableofcontents\fi

\section{INTRODUCTION}

A key obstacle in Bayesian statistics is to actually \emph{do} the
Bayesian inference. From a mathematical point of view, the inference
step is easy, transparent and defined by first principles: We simply
update prior beliefs about the unknown parameters with available
information in observed data, and obtain the posterior distribution
for the parameters. Based on the posterior, we can compute relevant
statistics for the parameters of interest, including marginal
distributions, means, variances, quantiles, credibility intervals,
etc. In practice, this is much easier said than done.

The introduction of simulation based inference, through the idea of
Markov chain Monte Carlo \citep{book44}, hit the statistical community
in the early 1990's and represented a major break-through in Bayesian
inference. MCMC provided a general recipe to generate samples from
posteriors by constructing a Markov chain with the target posterior as
the stationary distribution. This made it possible (in theory) to
extract and compute whatever one could wish for. Additional major
developments have paved the way for popular user-friendly MCMC-tools,
like \texttt{WinBUGS} \citep{tech23}, \texttt{JAGS} \citep{man2}, and
the new initiative \texttt{Stan} \citep{man3}, which uses Hamiltonian
Monte Carlo. Armed with these and similar tools, Bayesian statistics
has quickly grown in popularity and Bayesian statistics is now
well-represented in all the major research journals in all branches of
statistics.

In our opinion, however, from the point of view of applied users, the
impact of the Bayesian revolution has been less apparent. This is not
a statement about how Bayesian statistics itself is viewed by that
community, but about its rather ``cumbersome'' inference, which still
requires a lot of CPU -- and hence human time-- as well as tweaking of
simulation and model parameters to get it right. Re-running a lot of
alternative models gets even more cumbersome, making the iterative
process of model building in statistical analysis impossible
\citep[Sec.~1.1.4]{book77}. For this reason, simulation based
inference (and hence in most cases also Bayesian statistics) has too
often been avoided as being practically infeasible.

In this paper, we review a different take on doing Bayesian inference
that recently has facilitated the uptake of Bayesian modelling within
the community of applied users. The given approach is restricted to
the specific class of latent Gaussian models (LGMs) which, as will be
clear soon, includes a wide variety of commonly applied statistical
models making this restriction less limiting than it might appear at
first sight. The crucial point here is that we can derive
\emph{integrated nested Laplace approximation} (INLA methodology) for
LGMs, a deterministic approach to approximate Bayesian inference.
Performing inference within a reasonable time-frame, in most cases
INLA is both faster and more accurate than MCMC alternatives. Being
used to trading speed for accuracy this might seem like a
contradiction to most readers. The corresponding R-package (\RINLA,
see \url{www.r-inla.org}), has turned out to be very popular in
applied sciences and applied statistics, and has become a versatile
tool for quick and reliable Bayesian inference.

Recent examples of applications using the \RINLA package for
statistical analysis, include
disease mapping
\citep{schroedle-held-2011,schroedle-held-2011b,ugarte-etal-2014,ugarte-etal-2016,papoila-etal-2014,art592,art585},
age-period-cohort models \citep{riebler-held-2016},
evolution of the Ebola virus \citep{art593},
studies of relationship between access to housing, health and
well-being in cities \citep{art594},
study of the prevalence and correlates of intimate partner violence
against men in Africa \citep{art595},
search for evidence of gene expression heterosis \citep{art596},
analysis of traffic pollution and hospital admissions in London
\citep{art597},
early transcriptome changes in maize primary root tissues in response
to moderate water deficit conditions by RNA-Sequencing \citep{art598},
performance of inbred and hybrid genotypes in plant breeding and
genetics \citep{art599},
a study of Norwegian emergency wards \citep{art600},
effects of measurement errors \citep{art601,art561, muff-keller-2015},
network meta-analysis \citep{art602},
time-series analysis of genotyped human campylobacteriosis cases from
the Manawatu region of New Zealand \citep{art603},
modeling of parrotfish habitats \citep{art604},
Bayesian outbreak detection \citep{art605},
studies of long-term trends in the number of Monarch butterflies
\citep{art606},
long-term effects on hospital admission and mortality of road traffic
noise \citep{art607},
spatio-temporal dynamics of brain tumours \citep{art608},
ovarian cancer mortality \citep{art609},
the effect of preferential sampling on phylodynamic inference
\citep{art610},
analysis of the impact of climate change on abundance trends in
central Europe \citep{art611},
investigation of drinking patterns in US Counties from 2002 to 2012
\citep{art612},
resistance and resilience of terrestrial birds in drying climates
\citep{art613},
cluster analysis of population amyotrophic lateral sclerosis risk
\citep{art614},
malaria infection in Africa \citep{art615},
effects of fragmentation on infectious disease dynamics
\citep{art616},
soil-transmitted helminth infection in sub-Saharan Africa
\citep{art617},
analysis of the effect of malaria control on Plasmodium falciparum in
Africa between 2000 and 2015 \citep{art618},
adaptive prior weighting in generalized regression
\citep{held-sauter-2016},
analysis of hand, foot, and mouth disease surveillance data in China
\citep{art582},
estimate the biomass of anchovies in the coast of Per\'u
\citep{art549},
and many others.

We review the key components that make up INLA in \Sec{sec:background}
and in \Sec{sec:INLA} we combine these to outline why -- and in which
situations -- INLA works. In \Sec{sec:examples} we show some examples
of the use of \RINLA, and discuss some special features that expand
the class of models that \RINLA can be applied to. In
\Sec{sec:priors}, we discuss a specific challenge in Bayesian
methodology, and, in particular, reason why it is important to provide
better suggestions for default priors. We conclude with a general
discussion and outlook in \Sec{sec:discussion}.

\section{BACKGROUND ON THE KEY COMPONENTS}
\label{sec:background}

In this section, we review the key components of the INLA-approach to
approximate Bayesian inference. We introduce these concepts using a
top-down approach, starting with \emph{latent Gaussian models} (LGMs),
and what type of statistical models may be viewed as LGMs. We also
discuss the types of Gaussians/Gaussian-processes that are
computationally efficient within this formulation, and illustrate
Laplace approximation to perform integration -- a method that has been
around for a very long time yet proves to be a key ingredient in the
methodology we review here.

Due to the top-down structure of this text we occasionally have to
mention specific concepts before properly introducing and/or defining
them -- we ask the reader to bear with us in these cases.

\subsection{Latent Gaussian Models (LGMs)}

The concept of latent Gaussian models represents a very useful
abstraction subsuming a large class of statistical models, in the
sense that the task of statistical inference can be unified for the
entire class~\citep{art451}. This is obtained using a three-stage
hierarchical model formulation, in which observations $\mm{y}$ can be
assumed to be conditionally independent, given a latent Gaussian
random field $\mm{x}$ and hyperparameters $\mm{\theta_1}$,
\begin{equation*}\label{eq0a}
    \mm{y} \mid\mm{x},\mm{\theta}_1  \sim
    \prod_{i\in {\mathcal I}} \pi(y_i\mid x_i, \mm{\theta}_1).
\end{equation*}
The versatility of the model class relates to the specification of the
latent Gaussian field:
\begin{equation*}\label{eq0b}
    \mm{x} \mid\mm{\theta}_2   \sim
    \mathcal{N}\left(\mm{\mu}(\mm{\theta}_2),
      \mm{Q}^{-1}(\mm{\theta_2})\right)
\end{equation*}
which includes all random terms in a statistical model, describing the
underlying dependence structure of the data. The hyperparameters
$\mm{\theta}=(\mm{\theta}_1,\mm{\theta}_2)$, control the Gaussian
latent field and/or the likelihood for the data, and the posterior
reads
\begin{equation}\label{eq1}%
    \pi(\mm{x}, \mm{\theta} | \mm{y}) \propto
    \pi(\mm{\theta}) \;
    \pi(\mm{x} | \mm{\theta}) \;
    \prod_{i\in {\mathcal I}} \pi(y_i | x_i, \mm{\theta}).
\end{equation}
We make the following critical assumptions :
\begin{enumerate}
\item The number of hyperparameters $|\mm{\theta}|$ is \emph{small},
    typically $2$ to $5$, but not exceeding $20$.
\item The distribution of the latent field, $\mm{x}|\mm{\theta}$ is
    Gaussian and required to be a Gaussian Markov random field (GMRF)
    (or do be close to one) when the dimension $n$ is high ($10^{3}$
    to $10^{5}$).
\item The data \mm{y} are mutually conditionally independent of
    $\mm{x}$ and $\mm{\theta}$, implying that each observation $y_i$
    only depends on one component of the latent field, e.g.\ $x_i$.
    Most components of \mm{x} will not be observed.
\end{enumerate}
These assumptions are required both for computational reasons and to
ensure, with a high degree of certainty, that the approximations we
describe below are accurate.

\subsection{Additive Models}

Now, how do LGMs relate to other better-known statistical models?
Broadly speaking, they are an umbrella class generalising the large
number of related variants of ``additive'' and/or ``generalized''
(linear) models. For instance, interpreting the likelihood
$\pi(y_i|x_i, \mm{\theta})$, so that ``$y_i$ only depends on its
linear predictor $x_i$'', yields the generalized linear model setup.
We can interpret $\{x_i, i \in {\mathcal I}\}$ as $\eta_i$ (the linear
predictor), which itself is additive with respect to other effects,
\begin{equation}\label{eq3}%
    \eta_i = \mu + \sum_j \beta_j z_{ij} + \sum_k f_{k,j_k(i)}.
\end{equation}
Here, $\mu$ is the overall intercept and $\mm{z}$ are fixed covariates
with linear effects $\{\beta_j\}$. The difference between this
formulation and an ordinary generalized linear model are the terms
$\{f_{k}\}$, which are used to represent \emph{specific} Gaussian
processes. We label each $f_k$ as a \emph{model component}, in which
element $j$ contributes to the $i$th linear predictor. Examples of
model components $f_k$ include auto-regressive time-series models,
stochastic spline models and models for smoothing, measurement error
models, random effects models with different types of correlations,
spatial models etc. We assume that the model components are a-priori
independent, the fixed effects ($\mu, \mm{\beta})$ have a joint
Gaussian prior and that the fixed effects are a-priori independent of
the model components.

The key is now that the model formulation in \eref{eq3} and LGMs
relate to the same class of models when we assume Gaussian priors for
the intercept and the parameters of the fixed effects. The joint
distribution of
\begin{equation}\label{eq4}%
    \mm{x} = (\mm{\eta}, {\mu}, \mm{\beta}, \mm{f}_1, \mm{f}_2, \ldots)
\end{equation}
is then Gaussian, and also non-singular if we add a tiny noise term
in~\eref{eq3}. This yields the latent field \mm{x} in the hierarchical
LGM formulation. Clearly, $\text{dim}(\mm{x}) = n$ can easily get
large, as it equals the number of observations, plus the intercept(s)
and fixed effects, plus the sum of the dimension of all the model
components.

The hyperparameters $\mm{\theta}$ comprise the parameters of the
likelihood and the model components. A likelihood family and each
model component, typically has between zero and two hyperparameters.
These parameters often include some kind of variance, scale or
correlation parameters. Nicely, the number of hyperparameters is
typically small and further, does not depend on the dimension of the
latent field $n$ nor the number of observations. This is crucial for
computational efficiency, as even with a big dataset, the number of
hyperparameters remains constant and assumption 1. still holds.

\subsection{Gaussian Markov Random Fields (GMRFs)}

In practice, the latent field should not only be Gaussian, but should
also be a (sparse) Gaussian Markov random field (GMRF); see
\cite{book80,col27,col26} for an introduction to GMRFs. A GMRF
$\mm{x}$ is simply a Gaussian with additional conditional independence
properties, meaning that $x_i$ and $x_j$ are conditionally independent
given the remaining elements $\mm{x}_{-ij}$, for quite a few
$\{i,j\}$'s. The simplest non-trivial example is the first-order
auto-regressive model, $x_t = \phi x_{t-1} + \epsilon_t$,
$t=1, 2, \ldots, m$, having Gaussian innovations $\mm{\epsilon}$. For
this model, the correlation between $x_t$ and $x_s$ is $\phi^{|s-t|}$
and the resulting $m\times m$ covariance matrix is dense. However,
$x_s$ and $x_t$ are conditionally independent given $\mm{x}_{-st}$,
for all $|s-t|>1$. In the Gaussian case, a very useful consequence of
conditional independence is that this results in zeros for pairs of
conditionally independent values in the precision matrix (the inverse
of the covariance matrix). Considering GMRFs provides a huge
computational benefit, as calculations involving a dense $m \times m$
matrix are much more costly than when a sparse matrix is used. In the
auto-regressive example, the precision matrix is tridiagonal and can
be factorized in ${\mathcal O}(m)$ time, whereas we need
${\mathcal O}(m^{3})$ in the general dense case. Memory requirement is
also reduced, ${\mathcal O}(m)$ compared to ${\mathcal O}(m^{2})$,
which makes it much easier to run larger models. For models with a
spatial structure, the cost is ${\mathcal O}(m^{3/2})$ paired with a
${\mathcal O}(m\log(m))$ memory requirement. In general, the
computational cost depends on the actual sparsity pattern in the
precision matrix, hence it is hard to provide precise estimates.

\subsection{Additive Models and GMRFs}
In the construction of additive models including GMRFs the following
fact provides some of the ``magic'' that is exploited in INLA:
\begin{quote}
    \emph{The joint distribution for $\mm{x}$ in~\eref{eq4} is also a
        GMRF and its precision matrix consists of sums of the
        precision matrices of the fixed effects and the other model
        components.}
\end{quote}
We will see below that we need to form the joint distribution of the
latent field many times, as it depends on the hyperparameters
$\mm{\theta}$. Hence, it is essential that this can be done
efficiently avoiding computationally costly matrix operations. Being
able to simply treat the joint distribution as a GMRF with a precision
matrix that is easy to compute, is one of the key reasons why the
INLA-approach is so efficient. Also, the sparse structure of the
precision matrix boosts computationally efficiency, compared with
operations on dense matrices.

To illustrate more clearly what happens, let us consider the following
simple example,
\begin{equation}\label{eq5}%
    \eta_i = \mu + \beta z_i + f_{1j_1(i)} + f_{2j_2(i)} +
    \epsilon_i, \quad i=1, \ldots, n,
\end{equation}
where we have added a small amount of noise $\epsilon_i$. The two
model components $f_{1j_1(i)}$ and $f_{2j_2(i)}$ have sparse precision
matrices $\mm{Q}_1(\mm{\theta})$ and $\mm{Q}_2(\mm{\theta})$, of
dimension $m_1\times m_1$ and $m_2 \times m_2$, respectively. Let
$\tau_\mu$ and $\tau_\beta$ be the (fixed) prior precisions for $\mu$
and $\beta$. We can express~\eref{eq5} using matrices,
\begin{displaymath}
    \mm{\eta} = \mu\mm{1} + \beta \mm{z} + \mm{A}_1 \mm{f}_1 +
    \mm{A}_2 \mm{f}_2 + \mm{\epsilon}.
\end{displaymath}
Here, $\mm{A}_1$, and similarly for $\mm{A}_2$, is a $n\times m_1$
sparse matrix, which is zero except for exactly one $1$ in each row.
The joint precision matrix of
$(\mm{\eta}, \mm{f}_1, \mm{f}_2, \beta, \mu)$ is straight forward to
obtain by rewriting
\begin{eqnarray}\label{eq6}%
  &\exp\left(\right. -\frac{\tau_\epsilon}{2}
    \left(\mm{\eta} -\left(\mu\mm{1} + \beta \mm{z} + \mm{A}_1 \mm{f}_1 +
    \mm{A}_2 \mm{f}_2\right)\right)^{T}
    \left(\mm{\eta} -\left(\mu\mm{1} + \beta \mm{z} + \mm{A}_1 \mm{f}_1 +
    \mm{A}_2 \mm{f}_2\right)\right)
    \nonumber\\
  & \left.-\frac{\tau_\mu}{2} \mu^{2}
    -\frac{\tau_\beta}{2} \beta^{2}
    -\frac{1}{2} \mm{f}_1^{T} \mm{Q}_1(\mm{\theta}) \mm{f}_1
    -\frac{1}{2} \mm{f}_2^{T} \mm{Q}_2(\mm{\theta}) \mm{f}_2
    \right) \nonumber
\end{eqnarray}
into
\begin{displaymath}
    \exp\left( -\frac{1}{2} (\mm{\eta}, \mm{f}_1, \mm{f}_2, \beta, \mu)^{T}
      \mm{Q}_{\text{joint}}(\mm{\theta})
      (\mm{\eta}, \mm{f}_1, \mm{f}_2, \beta, \mu) \right)
\end{displaymath}
where
\begin{displaymath}
    \mm{Q}_{\text{joint}}(\mm{\theta}) =
    \begin{bmatrix}
        \tau_{\epsilon} \mm{I} & \tau_{\epsilon} \mm{A}_1 &
        \tau_{\epsilon} \mm{A}_2 & \tau_{\epsilon} \mm{I} \mm{z} &
        \tau_{\epsilon} \mm{I}\mm{1} \\
        & \mm{Q}_1(\mm{\theta}) + \tau_{\epsilon} \mm{A}_1
        \mm{A}_1^{T} & \tau_{\epsilon} \mm{A}_1 \mm{A}_2^{T} &
        \tau_{\epsilon} \mm{A}_1 \mm{z} & \tau_{\epsilon} \mm{A}_1
        \mm{1}\\
        && \mm{Q}_2(\mm{\theta}) + \tau_{\epsilon} \mm{A}_2
        \mm{A}_2^{T} & \tau_{\epsilon} \mm{A}_2 \mm{z} &
        \tau_{\epsilon} \mm{A}_2\mm{1} \\
        &\text{sym.}&& \tau_\beta + \tau_{\epsilon} \mm{z}^{T}\mm{z} &
        \tau_{\epsilon} \mm{z}^{T}\mm{1}\\
        &&&& \tau_\mu + \tau_{\epsilon} \mm{1}^{T}\mm{1}
    \end{bmatrix}.
\end{displaymath}
The dimension is $n + m_1 + m_2 + 2$. Concretely, the above-mentioned
``magic'' implies that the only matrices that need to be multiplied
are the \mm{A}-matrices, which are extremely sparse and contain only
one non-zero element in each row. These matrix products do not depend
on $\mm{\theta}$ and hence they only need to be computed once. The
joint precision matrix only depends on \mm{\theta} through
$\mm{Q}_1(\mm{\theta})$ and $\mm{Q}_2(\mm{\theta})$ and as
$\mm{\theta}$ change, the computational cost of re-computing
$\mm{Q}_\text{joint}(\mm{\theta})$ is negligible.

The sparsity of $\mm{Q}_{\text{joint}}(\mm{\theta})$ illustrates how
the additive structure of the model facilitates computational
efficiency. For simplicity, assume $n = m_1 = m_2$, and denote by
$e_1$ and $e_2$ the average number of non-zero elements in a row of
$\mm{Q}_1(\mm{\theta})$ and $\mm{Q}_2(\mm{\theta})$, respectively. An
upper bound for the number of non-zero terms in
$\mm{Q}_{\text{joint}}(\mm{\theta})$ is $n(19 + e_1 + e_2) +4$.
Approximately, this gives on average only $(19 + e_1 + e_2)/3$
non-zero elements for a row in $\mm{Q}_{\text{joint}}(\mm{\theta})$,
which is very sparse.

\subsection{Laplace Approximations}
\label{sec:laplace}

The Laplace approximation or method, is an old technique for the
approximation of integrals; see \cite[Ch.~3.3]{book123} for a general
introduction. The setting is as follows. The aim is to approximate the
integral,
\begin{displaymath}
    I_n = \int_x \exp(n f(x))\,dx
\end{displaymath}
as $n\rightarrow\infty$. Let $x_0$ be the point in which $f(x)$ has
its maximum, then
\begin{eqnarray}\label{eq7}%
  I_n &\approx& \int_x \exp\left(n\left(f(x_0) + \frac{1}{2}(x-x_0)^{2}
                f''(x_0)\right)\right)\, dx \\
      &=& {\exp(nf(x_0))}{\sqrt{\frac{2\pi}{-nf''(x_0)}}}
          = \widetilde{I}_n.
\end{eqnarray}
The idea is simple but powerful: Approximate the target with a
Gaussian, matching the mode and the curvature at the mode. By
interpreting $nf(x)$ as the sum of log-likelihoods and $x$ as the
unknown parameter, the Gaussian approximation will be exact as
$n\rightarrow\infty$, if the central limit theorem holds. The
extension to higher dimensional integrals, is immediate and the error
turns out to be
\begin{displaymath}
    I_n = \widetilde{I}_n \left(1 + {\mathcal O}(n^{-1})\right).
\end{displaymath}
This is a good result for two reasons. The error is \emph{relative}
and with rate $n^{-1}$, as opposed to an \emph{additive} error and a
rate $n^{-1/2}$, which are common in simulation-based inference.

The Laplace approximation used to be a key tool for doing
high-dimensional integration in pre-MCMC times, but quickly went out
of fashion when MCMC entered the stage. But how does it relate to what
we endeavour to do here? Lets assume that we would like to compute a
marginal distribution $\pi(\gamma_1)$ from a joint distribution
$\pi(\mm{\gamma})$
\begin{eqnarray}\label{eq8}%
  \pi(\gamma_1) &=& \frac{\pi(\mm{\gamma})}{\pi(\mm{\gamma}_{-1}|
                    \gamma_1)} \nonumber\\
                &\approx& \frac{\pi(\mm{\gamma})}{%
                          \pi_G(\mm{\gamma}_{-1};
                          \; \mm{\mu}(\gamma_1), \mm{Q}(\gamma_1))}
                          \Big|_{\mm{\gamma}_{-1} = \mm{\mu}(\gamma_1)},
\end{eqnarray}
where we have exploited the fact that we approximate
$\pi(\mm{\gamma}_{-1}|\gamma_1)$ with a Gaussian. In the context of
the LGMs we have $\mm{\gamma} = (\mm{x}, \mm{\theta})$. \cite{art367}
show that if $\pi(\mm{\gamma}) \propto \exp(nf_n(\mm{\gamma}))$, i.e.\
if $f_n(\mm{\gamma})$ is the average log likelihood, the relative
error of the \emph{normalized} approximation~\eref{eq8}, within a
${\mathcal O}(n^{-1/2})$ neighbourhood of the mode, is
${\mathcal O}(n^{-3/2})$. In other words, if we have $n$ replicated
data from the same parameters, $\mm{\gamma}$, we can compute posterior
marginals with a \emph{relative} error of ${\mathcal O}(n^{-3/2})$,
assuming the numerical error to be negligible. This is an extremely
positive result, but unfortunately the underlying assumptions usually
do not hold.
\begin{enumerate}
\item Instead of replicated data from the same model, we may have one
    replicate from one model (as is common in spatial statistics), or
    several observations from similar models.
\item The implicit assumption in the above result is also that
    $|\mm{\gamma}|$ is fixed as $n\rightarrow\infty$. However, there
    is only one realisation for each observation/location in the
    random effect(s) in the model, implying that $|\gamma|$ grows with
    $n$.
\end{enumerate}
Is it still possible to gain insight into when the Laplace
approximation would give good results, even if these assumptions do
not hold? First, let's replace \emph{replicated observations from the
    same model}, with several observations from \emph{similar} models
-- where we deliberately use the term ``similar'' in a loose sense. We
can borrow strength across variables that we \emph{a-priori} assume to
be similar, for example in smoothing over time or over space. In this
case, the resulting linear predictors for two observations could
differ in only one realisation of the random effect. In addition,
borrowing strength and smoothing can reduce the effect of the model
dimension growing with $n$, since the \emph{effective} dimension can
then grow much more slowly with $n$.

Another way to interpret the accuracy in computing posterior marginals
using Laplace approximations, is to not look at the error-rate but at
the implicit constant upfront. If the posterior is close to a Gaussian
density, the results will be more accurate compared to a density that
is very different from a Gaussian. This is similar to the convergence
for the central limit theorem where convergence is faster if relevant
properties such as uni-modality, symmetry and tail behaviour are
satisfied; see for example \cite{art619}. Similarly, in the context
here uni-modality is necessary since we approximate the integrand with
a Gaussian. Symmetry helps since the Gaussian distribution is
symmetric, while heavier tails will be missed by the Gaussian. For
example, assume
\begin{displaymath}
    \exp(n f_n(\mm{\gamma})) = \prod_i \text{Poisson}(y_i; \lambda =
    \exp(\gamma_1 + \gamma_2 z_i))
\end{displaymath}
with centred covariates $\mm{z}$. We then expect better accuracy for
$\pi(\gamma_1)$, having high counts compared with low counts. With
high counts, the Poisson distribution is approximately Gaussian and
almost symmetric. Low counts are more challenging, since the
likelihood for $y_i=0$ and $z_i=0$, is proportional to
$\exp(-\exp(\gamma_1))$, which has a maximum value at
$\gamma_1=-\infty$. The situation is similar for binomial data of size
$m$, where low values of $m$ are more challenging than high values of
$m$. Theoretical results for the current rather ``vague'' context are
difficult to obtain and constitute a largely unsolved problem; see for
example \cite{art408,art620,art621}.

Let us now discuss a simplistic, but realistic, model in two
dimensions $\mm{x} = (x_1, x_2)^{T}$, where
\begin{equation}\label{eq16}%
    \pi(\mm{x}) \propto \exp\left(-\frac{1}{2} \mm{x}^{T}
      \begin{bmatrix}
          1 & \rho \\ \rho & 1
      \end{bmatrix}
      \mm{x}\right)
    \prod_{i=1}^{2} \frac{\exp(cx_i)}{1+\exp(c x_i)}
\end{equation}
for a constant $c>0$ and $\rho\ge0$. This is the same functional form
as we get from two Bernoulli successes, using a logit-link. Using the
constant $c$ is an alternative to scaling the Gaussian part, and the
case where $\rho < 0$ is similar. The task now is to approximate
$\pi(x_1) = \pi(x_1, x_2)/\pi(x_2|x_1)$, using \eref{eq8}. Here, the
Gaussian approximation is indexed by $x_1$ and we use one Laplace
approximation for each value of $x_1$ . The likelihood term has a mode
at $(\infty, \infty)$, hence the posterior is a compromise between
this and the Gaussian prior centred at $(0,0)$.

\begin{figure}
    \centering
    \includegraphics[width=0.5\linewidth]{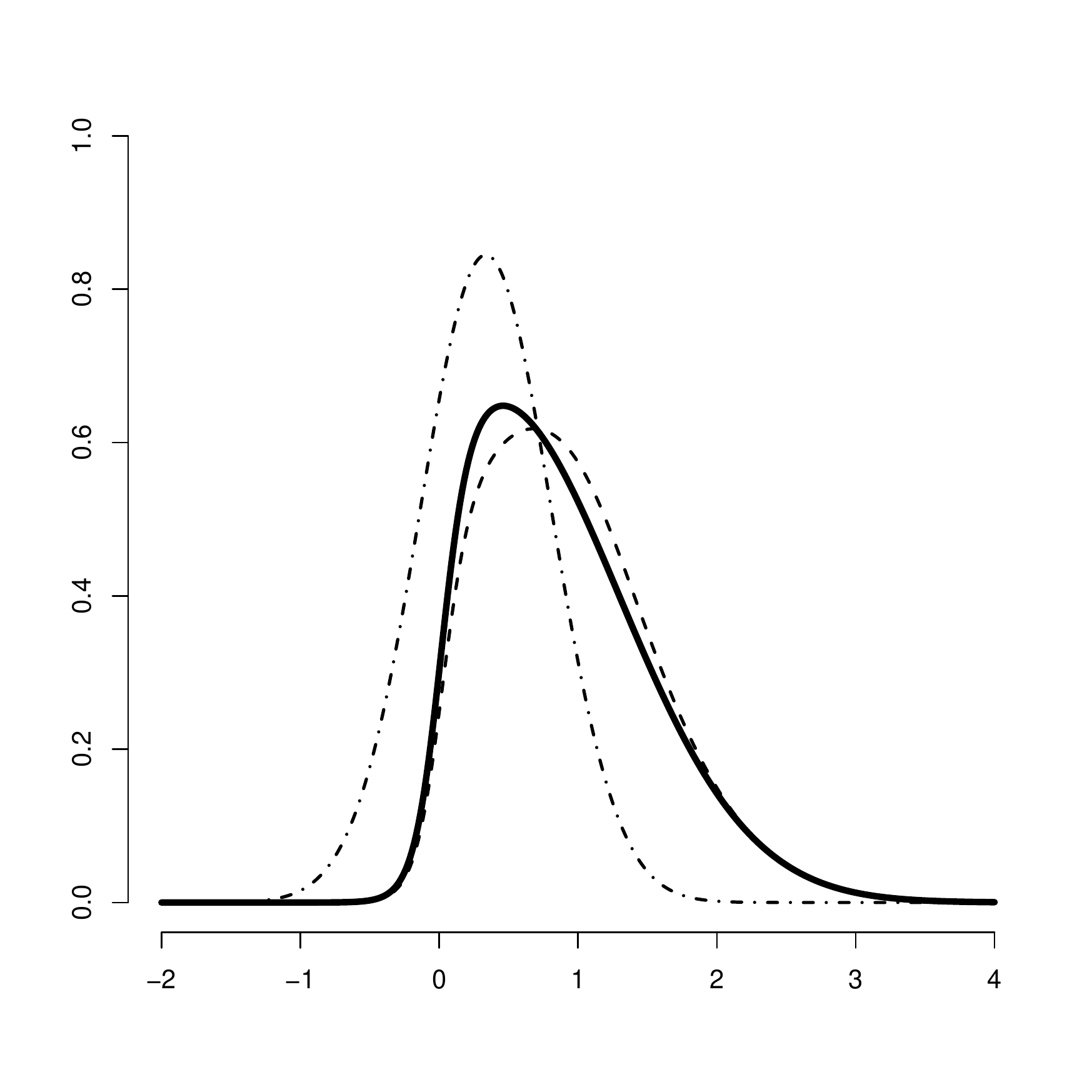}
    \caption{The true marginal (solid line), the Laplace approximation
        (dashed line) and the Gaussian approximation (dot-dashed
        line).}
    \label{fig1}
\end{figure}

We first demonstrate that even if the Gaussian approximation matching
the mode of $\pi(\mm{x})$ is not so good, the Laplace approximation
which uses a sequence of Gaussian approximations, can do much better.
Let $\rho=1/2$ and $c=10$ (which is an extreme value). The resulting
marginal for $x_1$ (solid), the Laplace approximation of it (dashed)
and Gaussian approximation (dot-dashed), are shown in \Fig{fig1}.
The Gaussian approximation fails both to locate the marginal correctly
and, of course, it also fails to capture the skewness that is present.
In spite of this, the \emph{sequence} of Gaussian approximations used
in the Laplace approximation performs much better and only seems to
run into slight trouble where the curvature of the likelihood changes
abruptly.

\begin{figure}
    \centering
    \begin{minipage}[b]{0.45\linewidth}
        \includegraphics[width=\linewidth]{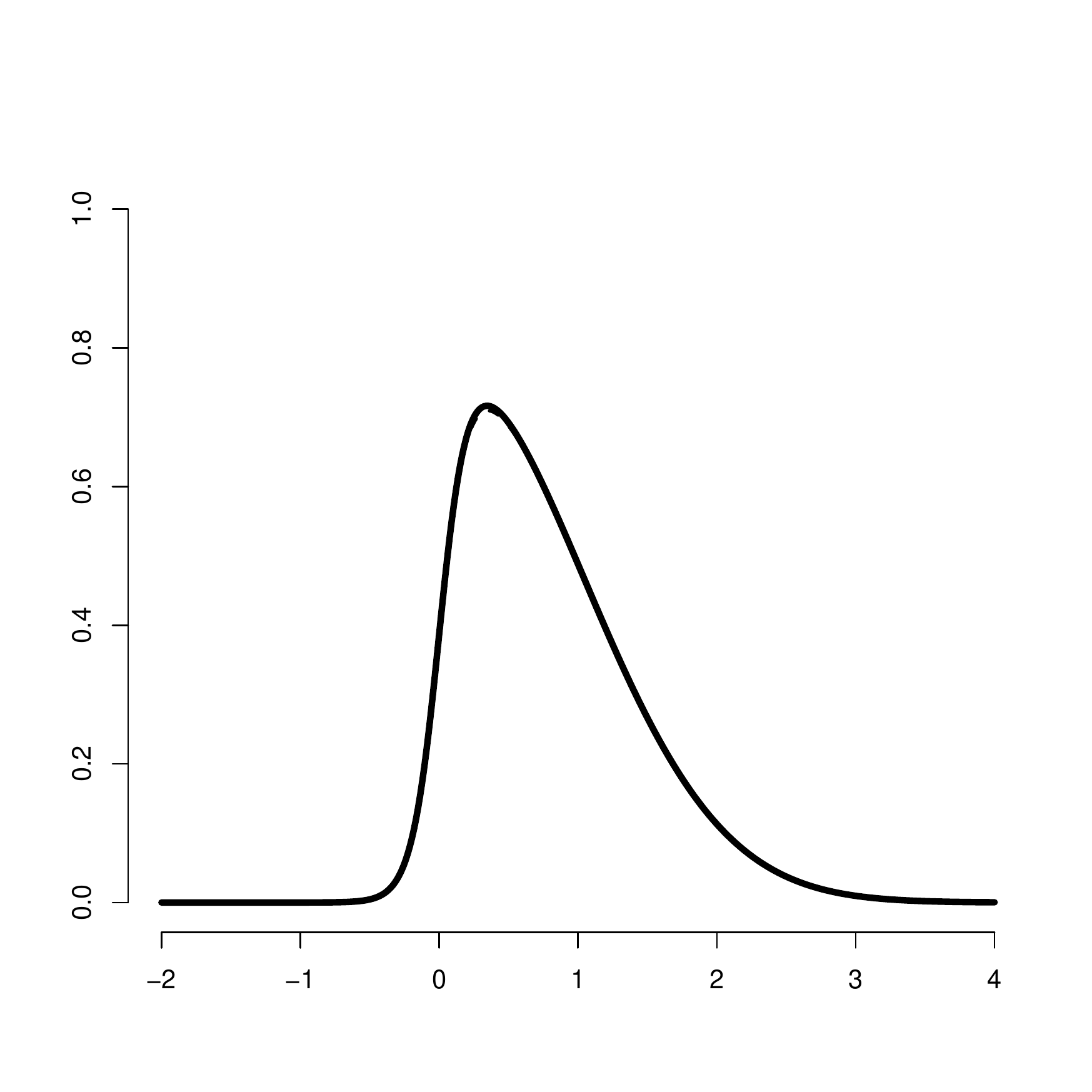}
        \vspace*{-1cm}\center{(a)}
    \end{minipage}
    \begin{minipage}[b]{0.45\linewidth}
        \includegraphics[width=\linewidth]{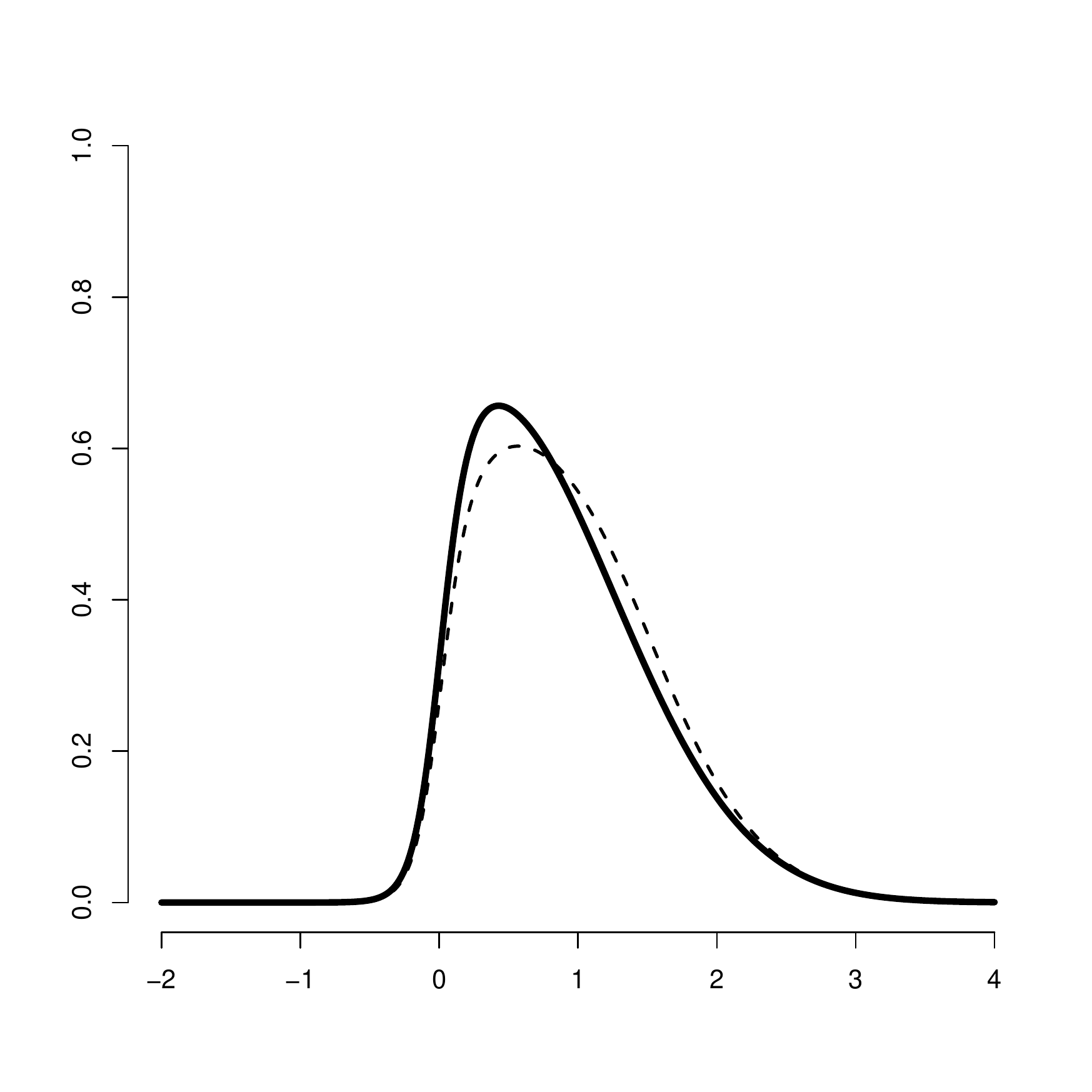}
        \vspace*{-1cm}\center{(b)}
    \end{minipage}
    \begin{minipage}[b]{0.45\linewidth}
        \includegraphics[width=\linewidth]{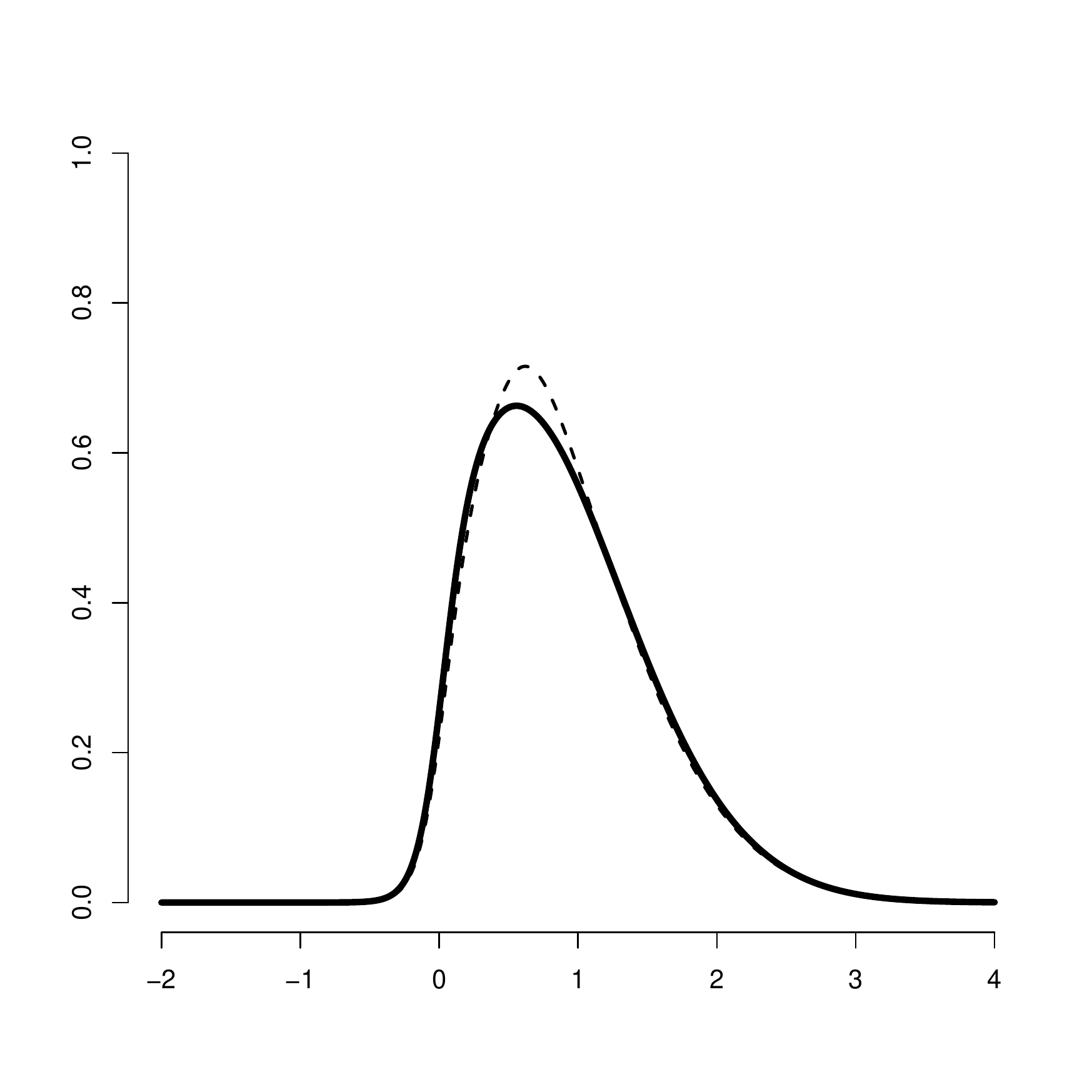}
        \vspace*{-1cm}\center{(c)}
    \end{minipage}
    \begin{minipage}[b]{0.45\linewidth}
        \includegraphics[width=\linewidth]{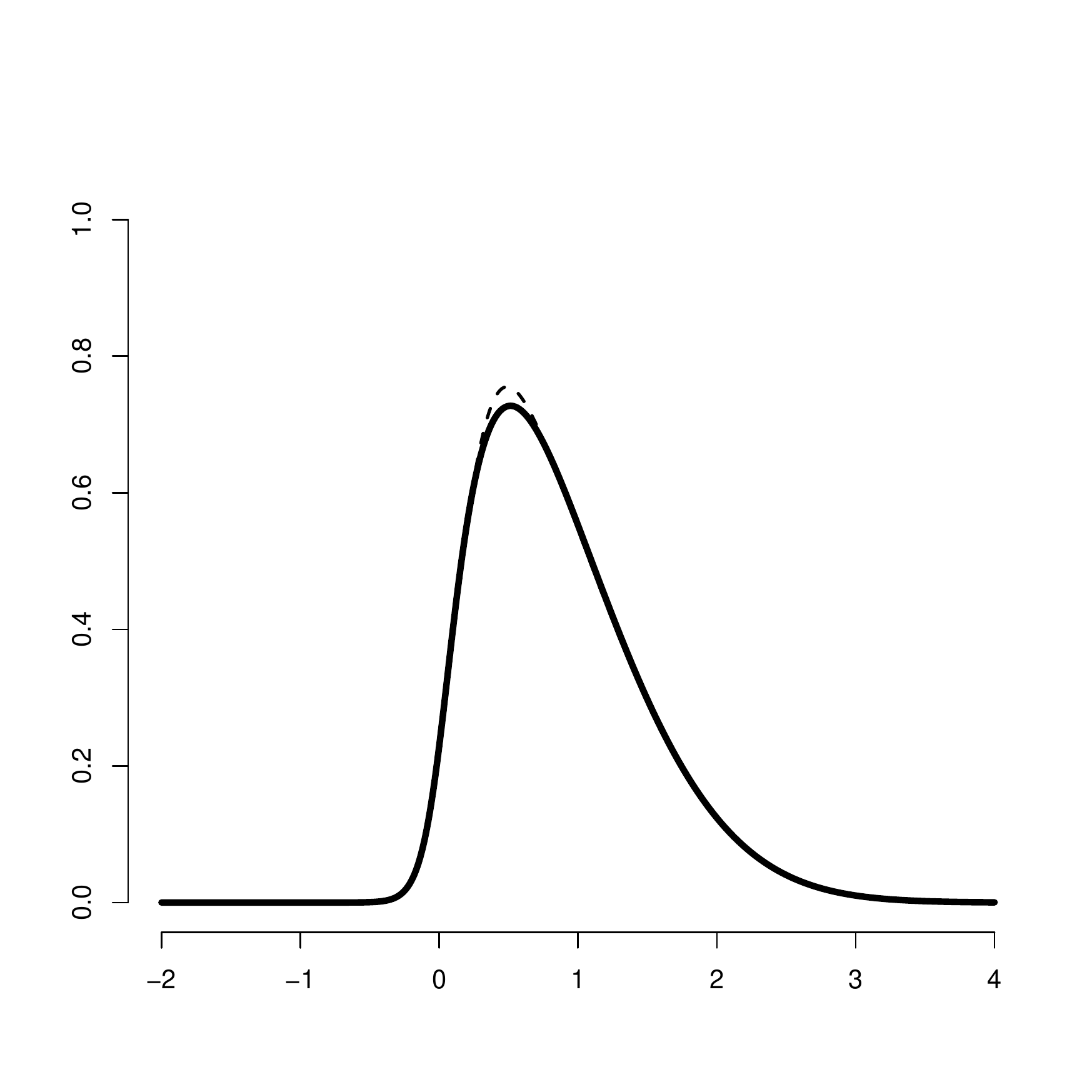}
        \vspace*{-1cm}\center{(d)}
    \end{minipage}
    \vspace*{1cm}
    \caption{The true marginal (solid line) and the Laplace
        approximation (dashed line), for $\rho = 0.05$ (a), $0.4$ (b),
        $0.8$ (c) and $0.95$ (d).}
    \label{fig2}
\end{figure}

An important feature of \eref{eq8} are its properties in the limiting
cases $\rho\rightarrow 0$ and $\rho\rightarrow 1$. When $\rho=0$,
$x_1$ and $x_2$ become independent and $\pi(x_2|x_1)$ does not depend
on $x_1$. Hence, \eref{eq8} is exact up to a numerical approximation
of the normalising constant. In the other limiting case,
$\rho\rightarrow 1$, $\pi(x_2|x_1)$ is the point-mass at $x_2=x_1$,
and \eref{eq8} is again exact up numerical error. This illustrates the
good property of \eref{eq8}, being exact in the two limiting cases of
weak and strong dependence, respectively. This indicates that the
approximation should not fail too badly for intermediate dependence.
\Fig{fig2} illustrates the Laplace approximation and the true
marginals, using $\rho=0.05, 0.4, 0.8$ and $0.95$, and $c=10$. For
$\rho=0.05$ (\Fig{fig2}a) and $\rho=0.95$ (\Fig{fig2}d), the
approximation is almost perfect, whereas the error is largest for
intermediate dependence where $\rho=0.4$ (\Fig{fig2}b) and $\rho=0.8$
(\Fig{fig2}c).

\section{Putting It All Together: INLA}
\label{sec:INLA}

With all the key components at hand, we now can put all these together
to illustrate how they are combined to from INLA. The main aim of
Bayesian inference is to approximate the posterior marginals
\begin{equation}\label{eq2}%
    \pi(\theta_j | \mm{y}), \quad j=1, \ldots, |\mm{\theta}|,
    \qquad
    \pi(x_i | \mm{y}), \quad i=1, \ldots, n.
\end{equation}
Our approach is tailored to the structure of LGMs, where
$|\mm{\theta}|$ is low-dimensional, $\mm{x}|\mm{\theta}$ is a GMRF and
the likelihood is conditional independent in the sense that $y_i$ only
depends on one $x_i$ and $\mm{\theta}$. From the discussion in
\Sec{sec:laplace}, we know that we should aim to apply Laplace
approximation only to near-Gaussian densities. For LGMs, it turns out
that we can reformulate our problem as series of subproblems that
allows us to use Laplace approximations on these. To illustrate the
general principal, consider an artificial model
\begin{displaymath}
    \eta_i = g(\beta) u_{j(i)},
\end{displaymath}
where $y_i | \eta_i \sim\text{Poisson}(\exp(\eta_i))$,
$i=1, \ldots, n$, $\beta\;\sim\;{\mathcal N}(0,1)$, $g(\cdot)$ is some
well-behaved monotone function, and
$\mm{u}\sim{\mathcal N}(\mm{0}, \mm{Q}^{-1})$. The index mapping
$j(i)$ is made such that the dimension of $\mm{u}$ is fixed and does
not depend on $n$, and all $u_j$s are observed roughly the same number
of times. Computation of the posterior marginals for $\beta$ and all
$u_j$ is problematic, since we have a product of a Gaussian and a
non-Gaussian (which is rather far from a Gaussian). Our strategy is to
break down the approximation into smaller subproblems and only apply
the Laplace approximation where the densities are almost Gaussian.
They key idea is to use conditioning, here on $\beta$. Then
\begin{equation}\label{eq9}%
    \pi(\beta|\mm{y}) \propto
    \pi(\beta) \int \prod_{i=1}^{n} \pi\left(y_i | \lambda_i
      = \exp\left(g(\beta)
        u_{j(i)}\right)\right) \times \pi(\mm{u}) \;d\mm{u}.
\end{equation}
The integral we need to approximate should be close to Gaussian, since
the integrand is a Poisson-count correction of a Gaussian prior. The
marginals for each $u_j$, can be expressed as
\begin{equation}\label{eq10}%
    \pi(u_j | \mm{y}) = \int
    \pi(u_j | \beta, \mm{y}) \times \pi(\beta
    | \mm{y}) \; d\beta.
\end{equation}
Note that we can compute the integral directly, since $\beta$ is
one-dimensional. Similar to \eref{eq9}, we have that
\begin{equation}\label{eq11}%
    \pi(\mm{u} | \beta, \mm{y}) \propto
    \prod_{i=1}^{n} \pi\left(y_i | \lambda_i = \exp\left(
        g(\beta) u_{j(i)}\right)\right) \times \pi(\mm{u}),
\end{equation}
which should be close to a Gaussian. Approximating
$\pi(u_j | \beta, \mm{y})$ involves approximation of the integral of
this density in one dimension less, since $u_j$ is fixed. Again, this
is close to Gaussian.

The key lesson learnt, is that we can break down the problem into
three sub-problems.
\begin{enumerate}
\item Approximate $\pi(\beta|\mm{y})$ using \eref{eq9}.
\item Approximate $\pi(u_j | \beta, \mm{y})$, for all $j$ and for all
    required values of $\beta$'s, from \eref{eq11}.
\item Compute $\pi(u_j|\mm{y})$ for all $j$ using the results from the
    two first steps, combined with numerical integration \eref{eq10}.
\end{enumerate}
The price we have to pay for taking this approach is increased
complexity; for example step 2 needs to be computed for all values of
$\beta$'s that are required. We also need to integrate out the
$\beta$'s in \eref{eq10}, numerically. If we remain undeterred by the
increased complexity, the benefit of this procedure is clear; we only
apply Laplace approximations to densities that are near-Gaussians,
replacing complex dependencies with conditioning and numerical
integration.

The big question is whether we can pursue the same principle for LGMs,
and whether we can make it computationally efficient by accepting
appropriate trade-offs that allow us to still be sufficiently exact.
The answer is \emph{Yes} in both cases. The strategy outlined above
can be applied to LGMs by replacing $\beta$ with $\mm{\theta}$, and
$\mm{u}$ with $\mm{x}$, and then deriving approximations to the
Laplace approximations and the numerical integration. The resulting
approximation is fast to compute, with little loss of accuracy. We
will now discuss the main ideas for each step -- skipping some
practical and computational details that are somewhat involved but
still relatively straight forward using ``every trick in the book''
for GMRFs.

\subsection{Approximating the Posterior Marginals for the
    Hyperparameters}
\label{sec:pmh}

Since the aim is to compute a posterior for each $\theta_j$, it is
tempting to use the Laplace approximation directly, which involves
approximating the distribution of
$(\mm{\theta}_{-j}, \mm{x})|(\mm{y}, \theta_j)$ with a Gaussian. Such
an approach will not be very successful, since the target is and will
not be very close to Gaussian; it will typically involve triplets like
$\tau x_i x_j$. Instead we can construct an approximation to
\begin{equation}\label{eq12}%
    \pi(\mm{\theta}|\mm{y}) \propto \frac{\pi(\mm{\theta}) \pi(\mm{x} |
        \mm{\theta}) \pi(\mm{y}|\mm{x}, \mm{\theta})}{%
        \pi(\mm{x} | \mm{\theta}, \mm{y})},
\end{equation}
in which the Laplace approximation requires a Gaussian approximation
of the denominator
\begin{eqnarray}\label{eq14}%
  \pi(\mm{x}|\mm{y}, \mm{\theta})
  &\propto&
            \exp\left(
            -\frac{1}{2} \mm{x}^{T}\mm{Q}(\mm{\theta}) \mm{x}
            + \sum_i \log\pi(y_i|x_i, \mm{\theta})
            \right) \\
  &=&
      \label{eq15}%
      (2\pi)^{-n/2} |\mm{P}(\theta)|^{1/2}
      \exp\left(
      -\frac{1}{2} (\mm{x} -\mm{\mu}(\mm{\theta}))^{T}
      \mm{P}(\mm{\theta})
      (\mm{x} -\mm{\mu}(\mm{\theta}))
      \right).
\end{eqnarray}
Here,
$\mm{P}(\mm{\theta}) = \mm{Q}(\theta) +
\text{diag}(\mm{c}(\mm{\theta}))$, while $\mm{\mu}(\mm{\theta})$ is
the location of the mode. The vector $\mm{c}(\mm{\theta})$ contains
the negative second derivatives of the log-likelihood at the mode,
with respect to $x_i$. There are two important aspects of \eref{eq15}.
\begin{enumerate}
\item It is a GMRF with respect to the same graph as from a model
    without observations $\mm{y}$, so computationally it does not cost
    anything to account for the observations since their impact is a
    shift in the mean and the diagonal of the precision matrix.
\item The approximation is likely to be quite accurate since the
    impact of conditioning on the observations, is only on the
    ``diagonal''; it shifts the mean, reduces the variance and might
    introduce some skewness into the marginals etc. Importantly, the
    observations do not change the Gaussian dependency structure
    through the terms $x_ix_j Q_{ij}(\mm{\theta})$, as these are
    untouched.
\end{enumerate}
Since $|\mm{\theta}|$ is of low dimension, we can derive marginals for
$\theta_j|\mm{y}$ directly from the approximation to
$\mm{\theta}|\mm{y}$. Thinking traditionally, this might be costly
since every new $\mm{\theta}$ would require an evaluation of
\eref{eq15} and the cost of numerical integration would still be
exponential in the dimension. Luckily, the problem is somewhat more
well-behaved, since the latent field $\mm{x}$ introduces quite some
uncertainty and more ``smooth'' behaviour on the $\mm{\theta}$
marginals.

In situations where the central limit theorem starts to kick in,
$\pi(\mm{\theta}|\mm{y})$ will be close to a Gaussian. We can improve
this approximation using variance-stabilising transformations of
$\mm{\theta}$, like using $\log(\text{precisions})$ instead of
precisions, the Fisher transform of correlations etc. Additionally, we
can use the Hessian at the mode to construct almost independent linear
combinations (or transformations) of $\mm{\theta}$. These
transformations really simplify the problem, as they tend to diminish
long tails and reduce skewness, which gives much simpler and
better-behaved posterior densities.

The task of finding a quick and reliable approach to deriving all the
marginal distributions from an approximation to the posterior density
\eref{eq12}, while keeping the number of evaluation points low, was a
serious challenge. We did not succeed on this until several years
after \cite{art451}, and after several failed attempts. It was hard to
beat the simplicity and stability of using the (Gaussian) marginals
derived from a Gaussian approximation at the mode. However, we needed
to do better as these Gaussian marginals were not sufficiently
accurate. The default approach used now is outlined in
\citet[Sec.~3.2]{art522}, and involves correction of local skewness
(in terms of difference in \emph{scale}) and an integration-free
method to approximate marginals from a skewness-corrected Gaussian.
How this is technically achieved is somewhat involved and we refer to
\cite{art522} for details. In our experience we now balance accuracy
and computational speed well, with an improvement over Gaussian
marginals while still being exact in the Gaussian limit.

In some situations, our approximation to \eref{eq12} can be a bit off.
This typically happens in cases with little smoothing and/or no
replications, for example when $\eta_i = \mu + \beta_z z_i + u_i$, for
a random-effect \mm{u}, and a binary likelihood
\citep{sauter-held-2016}. With vague priors model like this verge on
being improper. \cite{art587} discuss these cases and derive a
correction term which clearly improves the approximation to
$\pi(\mm{\theta}|\mm{y})$.

\subsection{Approximating the Posterior Marginals for the Latent
    Field}
\label{sec:pml}

We will now discuss how to approximate the posterior marginals for the
latent field. For linear predictors with no attached observations, the
posterior marginals are also the basis to derive the predictive
densities, as the linear predictor itself is a component of the latent
field. Similar to \eref{eq10}, we can express the posterior marginals
as
\begin{equation}\label{eq17}%
    \pi(x_i|\mm{y}) = \int \pi(x_i| \mm{\theta}, \mm{y}) \;
    \pi(\mm{\theta}| \mm{y}) \; d\mm{\theta},
\end{equation}
hence we are faced with two more challenges.
\begin{enumerate}
\item We need to integrate over $\pi(\mm{\theta}|\mm{y})$, but the
    computational cost of standard numerical integration is
    exponential in the dimension of $\mm{\theta}$. We have already
    ruled out such an approach in \Sec{sec:pmh}, since it was too
    costly computationally, except when the dimension is low.
\item We need to approximate $\pi(x_i|\mm{\theta}, \mm{y})$ for a
    subset of all $i=1, \ldots, n$, where $n$ can be (very) large,
    like in the range of $10^{3}$ to $10^{5}$. A standard application
    of the Laplace approximation, which involves location of the mode
    and factorisation of a $(n-1)\times(n-1)$ matrix many times for
    each $i$, will simply be too demanding.
\end{enumerate}
The key to success is to come up with efficient approximate solutions
for each of these problems.

\begin{figure}
    \centering
    \includegraphics[width=0.7\linewidth]{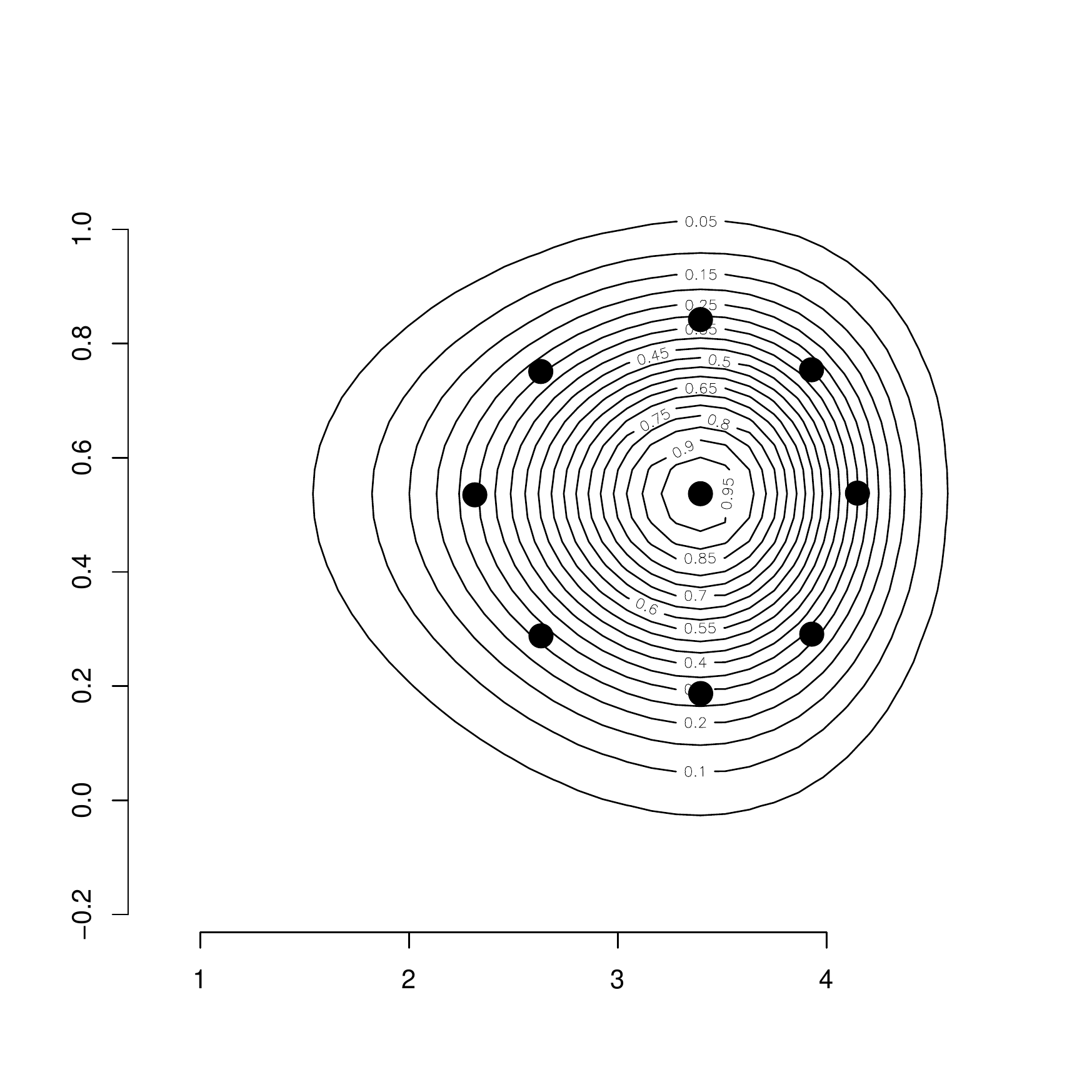}
    \caption{The contours of a posterior marginal for
        $(\theta_1, \theta_2)$ and the associated integration points
        (black dots).}
    \label{fig3}
\end{figure}

Classical numerical integration is only feasible in lower dimensions.
If we want to use $5$ integration points in each dimension, the cost
would be $5^{k}$ to cover all combinations in $k$ dimensions, which is
$125$ ($k=3$) and $625$ ($k=4$). Using only $3$ integration points in
each dimension, we get $81$ ($k=4$) and $729$ ($k=6$). This is close
to the practical limits. Beyond these limits we cannot aim to do
accurate integration, but should rather aim for something that is
better than avoiding the integration step, like an empirical Bayes
approach which just uses the mode. In dimensions $>2$, we borrow ideas
from central composite design \citep{art400} and use integration
points on a sphere around the centre; see \Fig{fig3} which illustrates
the procedure in dimension $2$ (even though we do not suggest using
this approach in dimension $1$ and $2$).
The integrand is approximately spherical (after rotation and scaling),
and the integration points will approximately be located on an
appropriate level set for the joint posterior of $\mm{\theta}$. We can
weight the spherical integration points equally, and determine the
relative weight with the central point requiring the correct
expectation of $\mm{\theta}^{T}\mm{\theta}$, if the posterior is
standard Gaussian \citep[Sec.~6.5]{art451}. It is our experience that
this approach balances computational costs and accuracy well, and it
is applied as the default integration scheme. More complex integration
schemes could be used with increased computational costs.

For the second challenge, we need to balance the need for improved
approximations beyond the Gaussian for $\pi(x_i|\mm{\theta}, \mm{y})$,
with the fact that we (potentially) need to do this $n$ times. Since
$n$ can be large, we cannot afford doing too heavy computations for
each $i$ to improve on the Gaussian approximations. The default
approach is to compute a Taylor expansion around the mode of the
Laplace approximation, which provides a linear and a cubic correction
term to the (standarized) Gaussian approximation,
\begin{equation}\label{eq13}%
    \log\pi(x_i | \mm{\theta}, \mm{y}) \approx -\frac{1}{2} x_i^{2} +
    b_i(\mm{\theta}) x_i + \frac{1}{6} c_i(\mm{\theta})x_i^{3}.
\end{equation}
We match a skew-Normal distribution~\citep{art414} to \eref{eq13},
such that the linear term provides a correction term for the mean,
while the cubic term provides a correction for skewness. This means
that we approximate \eref{eq17} with a mixture of skew-Normal
distributions. This approach, termed simplified Laplace approximation,
gives a very good trade-off between accuracy and computational speed.

Additional to posterior marginals, we can also provide estimates of
the deviance information criterion (DIC) \citep{art413},
Watanabe-Akaike information criterion (WAIC) \citep{art626,art627},
marginal likelihood and conditional predictive ordinates (CPO)
\citep{col28}. Other predictive criteria such as the ranked
probability score (RPS) or the Dawid-Sebastiani-Score (DSS)
\citep{art427} can also be derived in certain settings
\citep{art492,art498}. \cite{art531} discuss how the INLA-framework
can be extended to a class of near-Gaussian latent models.

\section{THE R-INLA PACKAGE: EXAMPLES}
\label{sec:examples}

The \RINLA package (see \texttt{www.r-inla.org}) provides an
implementation of the INLA-approach, including standard and
non-standard tools to define models based on the \texttt{formula}
concept in \texttt{R}. In this section, we present some examples of
basic usage and some special features of \RINLA.

\subsection{A Simple Example}
\label{sec:simple}

We first show the usage of the package through a simple simulated
example,
\begin{displaymath}
    \mm{y}|\mm{\eta} \;\sim\;\text{Poisson}(\exp(\mm{\eta}))
\end{displaymath}
where $\eta_i = \mu + \beta w_i + u_{j(i)}$, $i=1, \ldots, n$, $w$ are
covariates,
$\mm{u} \;\sim\; {\mathcal N}_m(\mm{0}, \tau^{-1} \mm{I})$, and $j(i)$
is a known mapping from $1:n$ to $1:m$. We generate data as follows
\begin{verbatim}
set.seed(123456L)
n = 50; m = 10
w = rnorm(n, sd = 1/3)
u = rnorm(m, sd = 1/4)
intercept = 0; beta = 1
idx = sample(1:m, n, replace = TRUE)
y = rpois(n, lambda = exp(intercept + beta * w + u[idx]))
\end{verbatim}
giving
\begin{verbatim}
> table(y, dnn=NULL)
 0  1  2  3  5
17 18  9  5  1
\end{verbatim}
We use \RINLA to do the inference for this model, by
\begin{verbatim}
library(INLA)
my.data = data.frame(y, w, idx)
formula = y ~ 1 + w + f(idx, model="iid"),
r = inla(formula, data = my.data, family = "poisson")
\end{verbatim}
The \texttt{formula} defines how the response depends on covariates,
as usual, but the term \texttt{f(idx, model="iid")} is new. It
corresponds to the function $f$ that we have met above in \eref{eq3},
one of many implemented GMRF model components. The \texttt{iid} term
refers to the ${\mathcal N}(\mm{0}, \tau^{-1}\mm{I})$ model, and
\texttt{idx} is an index that specifies which elements of the model
component go into the linear predictor.

\begin{figure}
    \centering
    \begin{minipage}[b]{0.49\linewidth}
        \includegraphics[width=\linewidth]{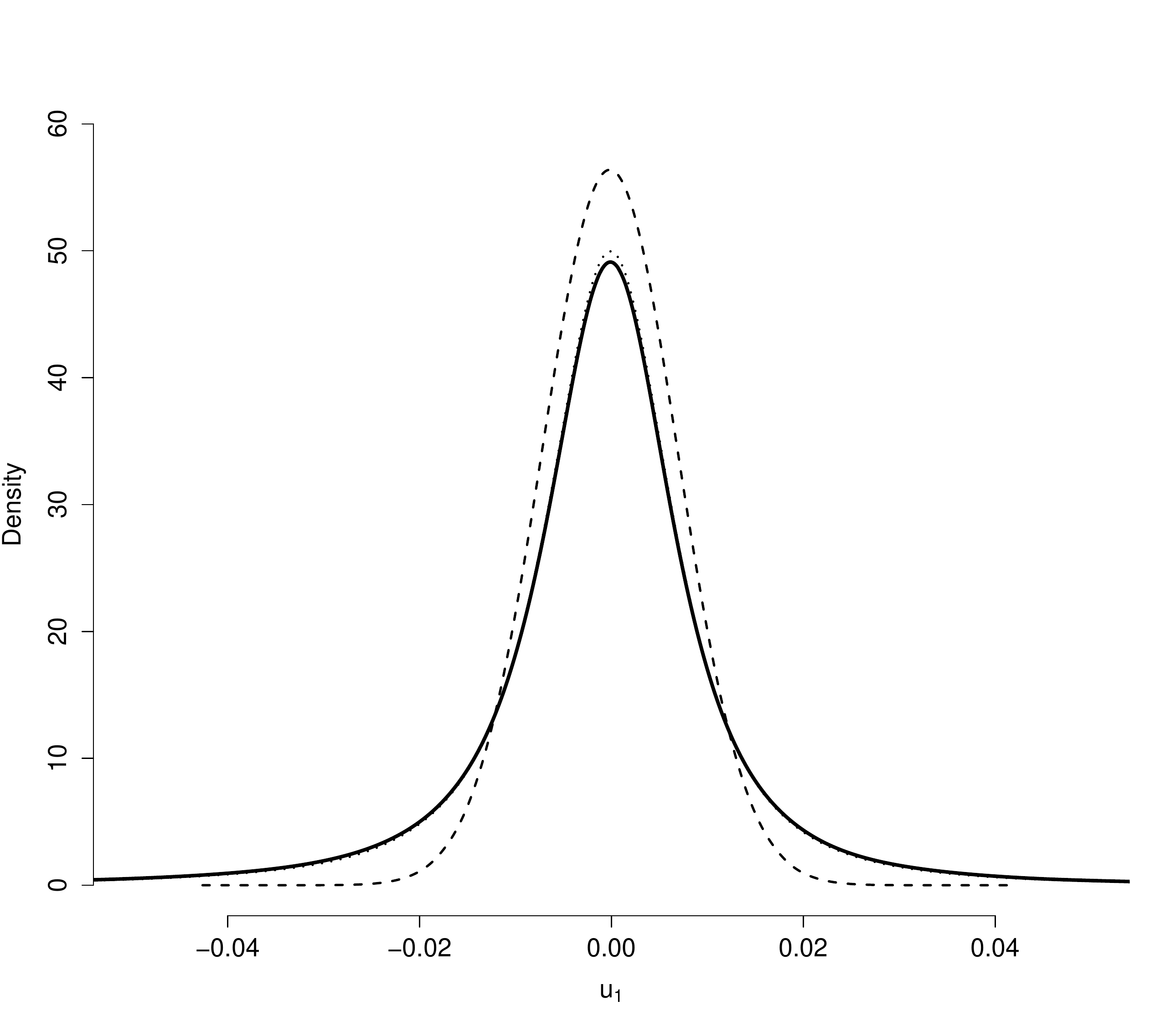}
        \vspace*{-1cm}\center{(a)}
    \end{minipage}
    \begin{minipage}[b]{0.49\linewidth}
        \includegraphics[width=\linewidth]{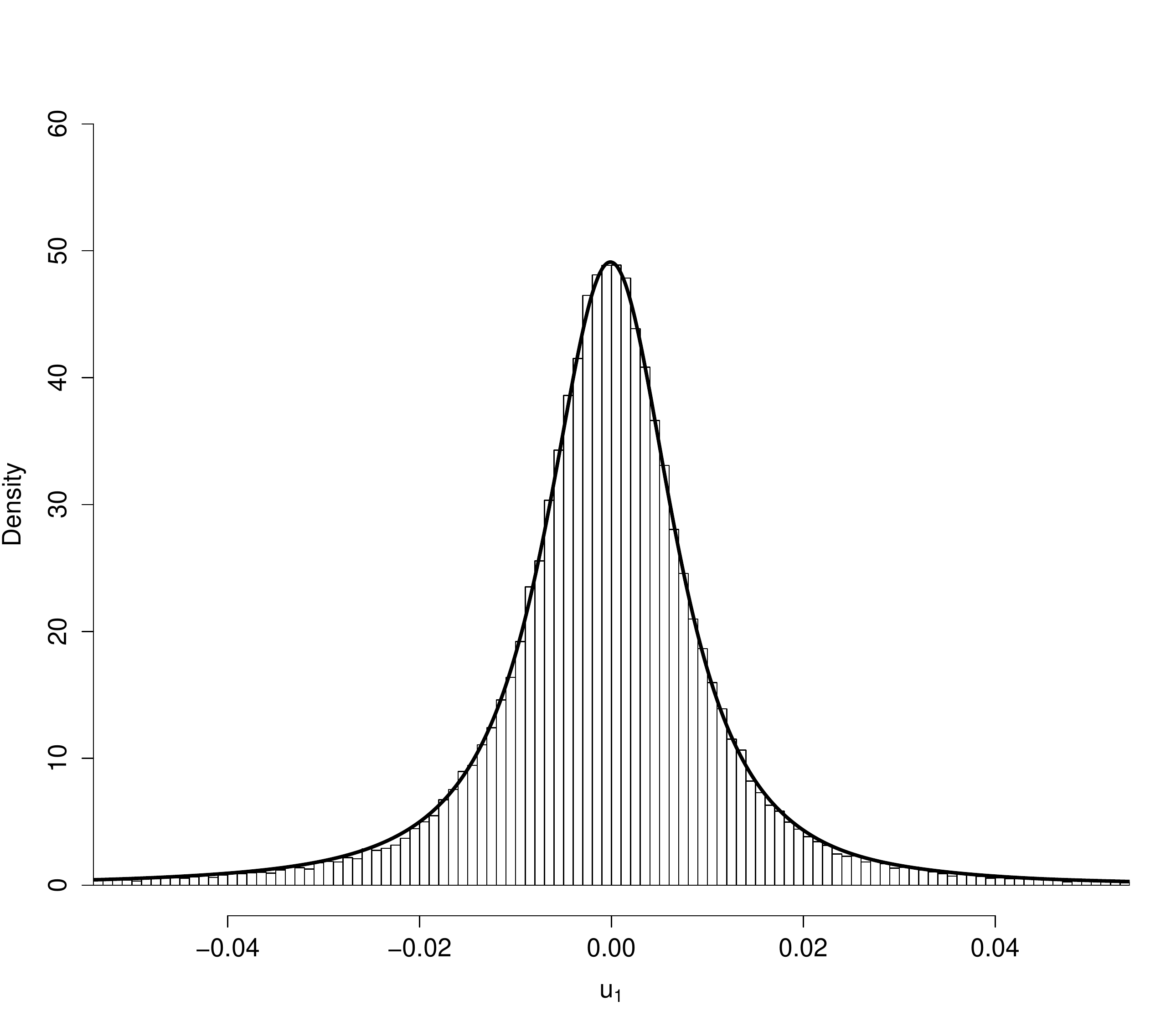}
        \vspace*{-1cm}\center{(b)}
    \end{minipage}
    \vspace*{5mm}
    \caption{Panel (a) shows the default estimate (simplified Laplace
        approximation) of the posterior marginal for $u_1$ (solid), a
        simplified estimate, i.e. the Gaussian approximation, (dashed)
        and the best possible Laplace approximation (dotted). Panel
        (b) shows the histogram of $u_1$ using $10^{5}$ samples
        produced using \texttt{JAGS}, together with the simplified
        Laplace approximation from (a).}
    \label{fig4}
\end{figure}

\Fig{fig4}a shows three estimates of the posterior marginal of $u_1$.
The solid line is the default estimate, the simplified Laplace
approximation, as outlined in \Sec{sec:INLA} (and with the
\texttt{R}-commands given above). The dashed line is the simpler
Gaussian approximation which avoids integration over $\mm{\theta}$,
\begin{verbatim}
r.ga = inla(formula, data = my.data, family = "poisson",
            control.inla = list(strategy = "gaussian", int.strategy = "eb"))
\end{verbatim}
The dotted line represents the (almost) true Laplace approximations
and accurate integration over $\mm{\theta}$, and is the best
approximation we can provide with the current software,
\begin{verbatim}
r.la = inla(formula, data = my.data, family = "poisson",
            control.inla = list(strategy = "laplace",
                int.strategy = "grid", dz=0.1, diff.logdens=20))
\end{verbatim}
It is hard to see as it almost entirely covered by the solid line,
meaning that our mixture of skew-Normals is very close to being exact
in this example. We also note that by integrating out $\mm{\theta}$,
the uncertainty increases, as it should. To compare the approximations
with a simulation based approach, \Fig{fig4}b shows the corresponding
histogram for $10^{5}$ samples using \texttt{JAGS}, together with the
default estimate from \Fig{fig4}a. The fit is quite accurate. The CPU
time used by \RINLA with default options, was about $0.16$ seconds on
a standard laptop where $2/3$ of this time was used for
administration.

\subsection{A Less Simple Example Including Measurement Error}
\label{sec:less.simple}

We continue with a measurement error extension of the previous
example, assuming that the covariate $\mm{w}$ is only observed
indirectly through $\mm{z}$, where
\begin{displaymath}
    z_i|\ldots \;\sim\; \texttt{Binomial}\left(m,
      \text{prob} = \frac{1}{1+\exp(-(\gamma + w_i))}\right), \quad i=1,
    \ldots, n,
\end{displaymath}
with intercept $\gamma$. In this case, the model needs to be specified
using two likelihoods and also a special feature called \texttt{copy}.
Each observation can have its own type of likelihood (i.e.\ family),
which is coded using a matrix (or list) of observations, where each
``column'' represents one family. A linear predictor can only be
associated with one observation. The \texttt{copy} feature allows us
to have additional identical copies of the \emph{same} model component
in the formula, and we have the option to scale it as well. An index
\texttt{NA} is used to indicate if there is no contribution to the
linear predictor and this is used to zero-out contributions from model
components. This is done in the code below:
\begin{verbatim}
## generate observations that we observe for 'w'
m = 2
z = rbinom(n, size = m, prob = 1/(1+exp(-(0 + w))))
## create the response. since we have two families, poisson and
## binomial, we use a matrix, one column for each family
Y = matrix(NA, 2*n, 2)
Y[1:n    , 1] = y
Y[n + 1:n, 2] = z
## we need one intercept for each family. this is an easy way to achive that
Intercept = as.factor(rep(1:2, each=n))
## say that we have 'beta*w' only for 'y' and 'w' only for 'z'. the formula
## defines the joint model for both the observations, 'y' and 'z'
NAs  = rep(NA, n)
idx  = c(NAs, 1:n)
idxx = c(1:n, NAs)
formula2 = Y ~ -1 + Intercept + f(idx, model="iid") +
               f(idxx, copy="idx", hyper = list(beta = list(fixed = FALSE)))
## need to use a 'list' since 'Y' is a matrix
my.data2 = list(Y=Y, Intercept = Intercept, idx = idx,  idxx = idxx)
## we need to define two families and give the 'size' for the binomial
r2 = inla(formula2, data = my.data2, family = c("poisson", "binomial"),
          Ntrials = c(NAs, rep(m, n)))
\end{verbatim}
We refer to \cite{art561} for more details on measurement error models
using INLA, and to the specific latent Gaussian models termed {\tt
    mec} and {\tt meb} that are available in {\tt R-INLA} to
facilitate the implementation of classical error models and Berkson
error models, respectively.

\subsection{A Spatial Example}
\label{sec:spatial}

The \RINLA package has extensive support for spatial Gaussian models,
including intrinsic GMRF models on regions (often called ``CAR''
models, \citep[Ch.~5.2]{book124}), and a subclass of continuously
indexed Gaussian field models. Of particular interest are Gaussian
fields derived from stochastic partial differential equations (SPDEs).
The simplest cases are Mat\'ern fields in dimension $d$, which can be
described as the solution to
\begin{equation}\label{eq18}%
    (\kappa^{2} - \Delta)^{\alpha/2} (\tau {x}(\mm{s})) =
    {\mathcal W}(\mm{s}),
\end{equation}
where $\Delta$ is the Laplacian, $\kappa>0$ is the spatial scale
parameter, $\alpha$ controls the smoothness, $\tau$ controls the
variance, and ${\mathcal W}(\mm{s})$ is a Gaussian spatial white noise
process. \cite{art246,art455} shows that its solution is a Gaussian
field with a Mat\'ern covariance function having smoothness
$\nu = \alpha-d/2$. The smoothness is usually kept fixed based on
prior knowledge of the underlying process. A formulation of Mat\'ern
fields as solutions to \eref{eq18} might seem unnecessarily
complicated, since we already know the solution. However,
\cite{art500} showed that by using a finite basis-function
representation of the continuously indexed solution, one can derive
(in analogy to the well known \emph{Finite Element Method}) a local
representation with Markov properties. This means that the joint
distribution for the weights in the basis-function expansion is a
GMRF, and the distribution follows directly from the basis functions
and the triangulation of space. The main implication of this result is
that it allows us to continue to think about and interpret the model
using marginal properties like covariances, but at the same time we
can do fast computations since the Markov properties make the
precision matrix very sparse. It also allows us to add this component
in the \RINLA framework, like any other GMRF model-component.

The dual interpretation of Mat\'ern fields, both using covariances and
also using its Markov properties, is very convenient both from a
computational but also from a statistical modeling point of view
\citep{art512,art508,art527}. The same ideas also apply to
non-stationary Gaussian fields using non-homogeneous versions of an
appropriate SPDE \citep{art500,art529,art581,art532}, Gaussian fields
that treats land as a barrier to spatial correlation \citep{tech125},
multivariate random fields \citep{art591}, log-Gaussian Cox processes
\citep{art583}, and in the near future also to non-separable
space-time models.

We end this section with a simple example of spatial survival analysis
taken from \cite{art458}, studying spatial variation in leukaemia
survival data in north-west England in the period 1982--1998. The
focus of the example is to see how and how easily, the spatial model
integrates into the model definition \citep{art494}. We therefore omit
further details about the dataset and refer to the original article.

\begin{figure}
    \centering
    \begin{minipage}[b]{0.49\linewidth}
        \includegraphics[width=\linewidth]{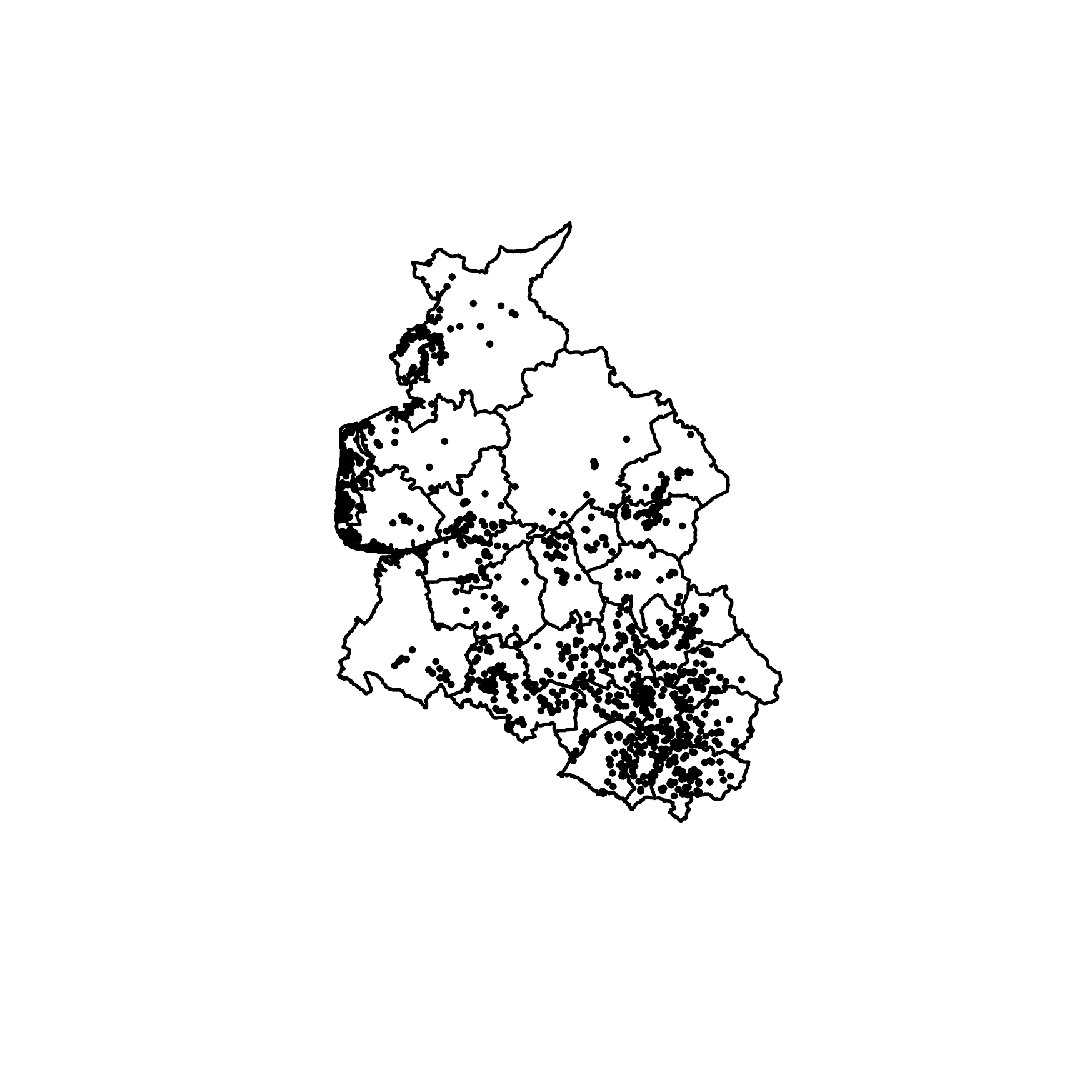}
        \vspace*{-1cm}\center{(a)}
    \end{minipage}
    \begin{minipage}[b]{0.49\linewidth}
        \includegraphics[width=\linewidth]{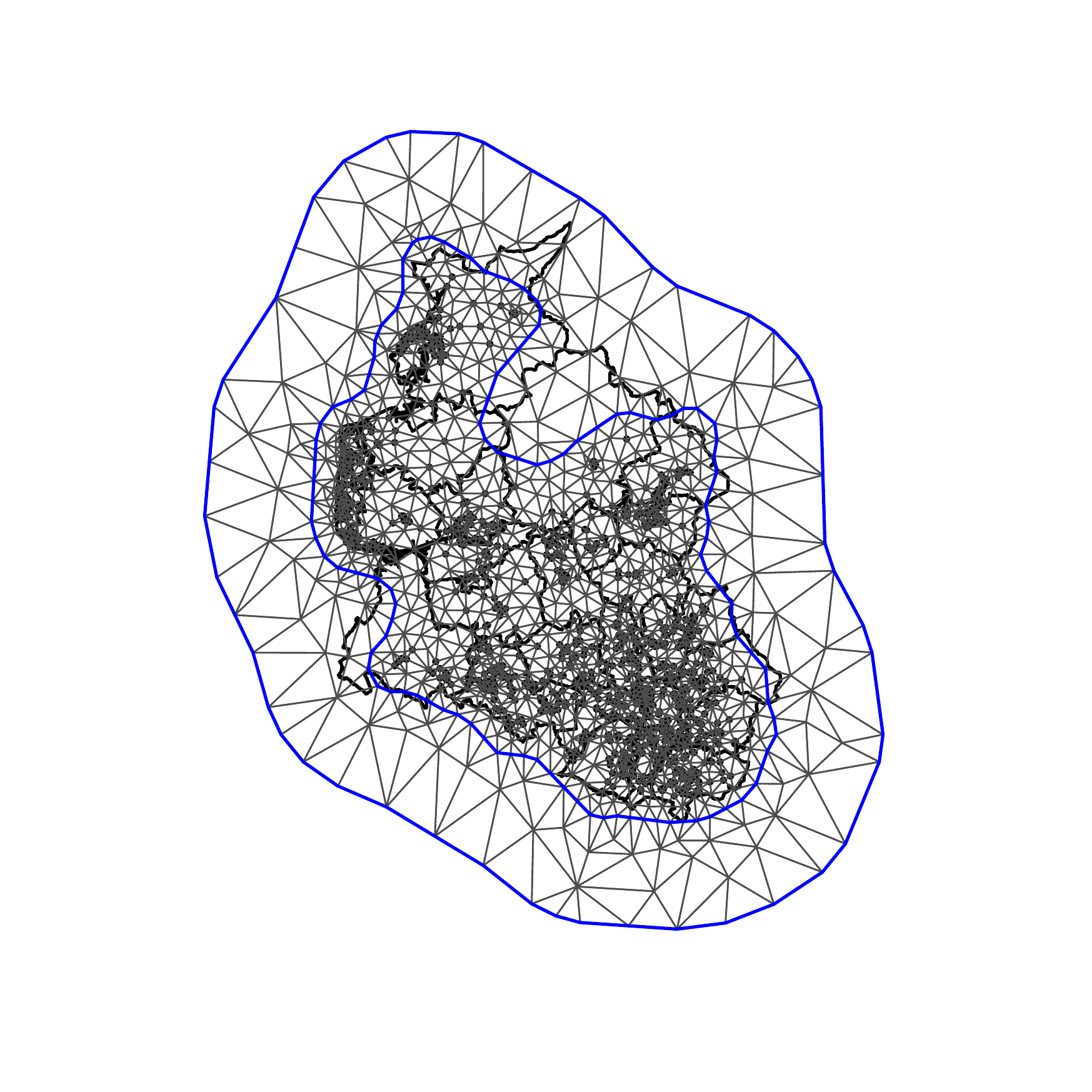}
        \vspace*{-1cm}\center{(b)}
    \end{minipage}
    \vspace*{5mm}
    \caption{Panel (a) shows the area of north-west England for the
        leukaemia study, where the (post-code) locations of the events
        are shown as dots. Panel (b) overlays the mesh used for the
        SPDE model.}
    \label{fig5}
\end{figure}

First, we need to load the data and create the mesh, i.e.\ a
triangulation of the area of interest to represent the finite
dimensional approximation to \eref{eq18}.
\begin{verbatim}
library(INLA)
data(Leuk)
loc <- cbind(Leuk$xcoord, Leuk$ycoord)
bnd1 <- inla.nonconvex.hull(loc, convex=0.05)
bnd2 <- inla.nonconvex.hull(loc, convex=0.25)
mesh <- inla.mesh.2d(loc, boundary=list(bnd1, bnd2),
                     max.edge=c(0.05, 0.2), cutoff=0.005)
\end{verbatim}
\Fig{fig5}a displays the study area and the locations of the events,
while \Fig{fig5}b shows the associated mesh with respect to which we
define the SPDE model. We use an additional rougher mesh to reduce
boundary effects.
The next step is to create a mapping matrix from the mesh onto the
locations where the data are observed. Then we define the SPDE model,
to define the statistical model including covariates like sex, age,
white blood-cell counts (wbc) and the Townsend deprivation index
(tpi), and to call a book-keeping function which keeps the indices in
correct order. Finally, we call \texttt{inla()} to do the analysis,
assuming a Weibull likelihood. Note that application of a Cox
proportional hazard model will give similar results.

\begin{verbatim}
A <- inla.spde.make.A(mesh, loc)
spde <- inla.spde2.matern(mesh, alpha=2) ## alpha=2 is the default choice
formula <- inla.surv(time, cens) ~ 0 + a0 + sex + age + wbc + tpi +
                     f(spatial, model=spde)
stk <- inla.stack(data=list(time=Leuk$time, cens=Leuk$cens), A=list(A, 1),
                  effect=list(list(spatial=1:spde$n.spde),
                  data.frame(a0=1, Leuk[,-c(1:4)])))
r <- inla(formula, family="weibull", data=inla.stack.data(stk),
          control.predictor=list(A=inla.stack.A(stk)))
\end{verbatim}

\begin{figure}
    \centering
    \includegraphics[width=0.5\linewidth]{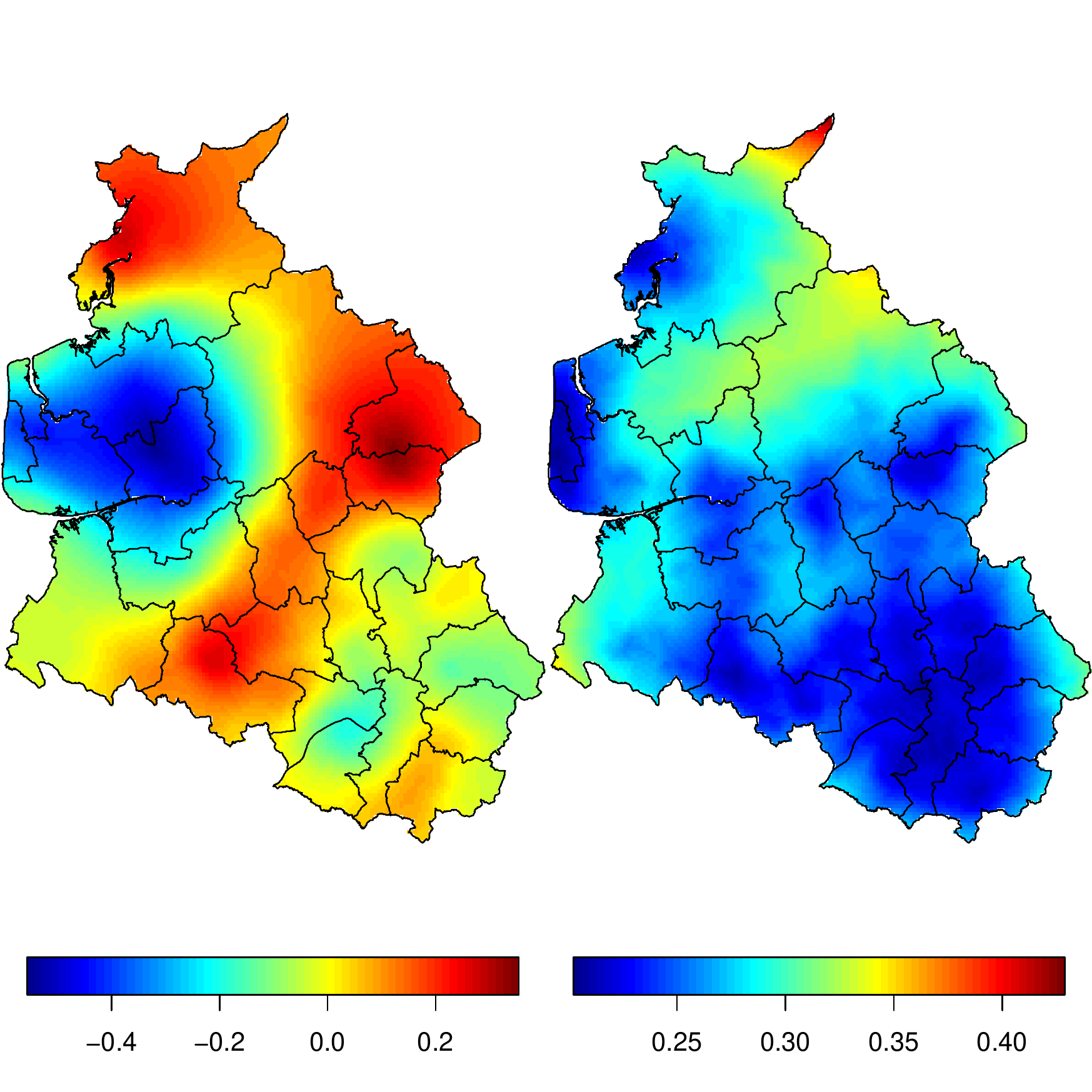}
    \caption{The spatial effect in the model (left: mean, right:
        standard deviation).}
    \label{fig6}
\end{figure}

\Fig{fig6}a shows the estimated spatial effect, with the posterior
mean (left), and posterior standard deviation (right).

\subsection{Special Features}

In addition to standard analyses, the \RINLA package also contains
non-standard features that really boost the complexity of models that
can be specified and analysed. Here, we give a short summary of these,
for more details see \cite{art522}.
\begin{description}
\item[replicate] Each model component given as a \texttt{f()}-term can
    be replicated, creating \texttt{nrep} iid replications with shared
    hyperparameters. For example,
\begin{verbatim}
f(time, model="ar1", replicate=person)
\end{verbatim}
    defines one AR(1) model for each person sharing the same
    hyperparameters.
\item[group] Each model component given as a \texttt{f()}-term, can be
    grouped, creating \texttt{ngroup} dependent replications with a
    separable correlation structure. To create a separable space-time
    model, with an AR(1) dependency in time, we can specify
\begin{verbatim}
f(space, model=spde, group=time, control.group = list(model = "ar1"))
\end{verbatim}
    \cite{art492} used grouped smoothing priors in \RINLA to impute
    missing mortality rates for a specific country by taking advantage
    from similar countries where these data are available. The authors
    provide the corresponding R-code in the supplementary material. We
    can both group and replicate model components.
\item[A-matrix] We can create a second layer of linear predictors
    where \mm{\eta} is defined by the \texttt{formula}, but where
    $\mm{\eta}^{*} = \mm{A}\mm{\eta}$ is connected to the
    observations. Here, \mm{A} is a constant (sparse) matrix; see the
    above spatial example.
\item[Linear combinations] We can also compute posterior marginals of
    $\mm{v} = \mm{B}\mm{x}$ where \mm{x} is the latent field and
    $\mm{B}$ is a fixed matrix. This could for example be
    $\beta_1 - \beta_2$ for two fixed effects, or any other linear
    combinations. Here is an example computing the posterior for the
    difference between two linear effects, $\beta_u - \beta_{v}$
\begin{verbatim}
lc = inla.make.lincomb(u=1, v=-1)
r = inla(y ~ u + v, data = d, lincomb = lc)
\end{verbatim}
\item[Remote server] It is easy to set up a remote MacOSX/Linux server
    to host the computations while doing the \texttt{R}-work at your
    local laptop. The job can be submitted and the results can be
    retrieved later, or we can use it interactively. This is a very
    useful feature for larger models. It also ensures that
    computational servers will in fact be used, since we can work in a
    local \texttt{R}-session but use a remote server for the
    computations. Here is an example running the computations on a
    remote server
\begin{verbatim}
r = inla(formula, family, data = data, inla.call = "remote")
\end{verbatim}
    To submit a job we specify
\begin{verbatim}
r = inla(formula, family, data = data, inla.call = "submit")
\end{verbatim}
    and we can check the status and retrieve the results when the
    computations are done, by
\begin{verbatim}
inla.qstat(r)
r = inla.qget(r)
\end{verbatim}
\item[R-support] Although the core inla-program is written in
    \texttt{C}, it is possible to pass a user-defined latent model
    component written in \texttt{R}, and use that as any other latent
    model component. The \texttt{R}-code will be evaluated within the
    \texttt{C}-program. This is very useful for more specialised model
    components or re-parameterisations of existing ones, even though
    it will run slower than a proper implementation in \texttt{C}. As
    a simple example, the code below implements the model component
    \texttt{iid}, which is just independent Gaussian random effects
    ${\mathcal N}_n(\mm{0}, (\tau\mm{I})^{-1})$. The skeleton of the
    function is predefined, and must return the graph, the
    \mm{Q}-matrix, initial values, the mean, the log normalising
    constant and the log prior for the hyperparameter.
\begin{verbatim}
iid.model = function(cmd = c("graph", "Q", "mu", "initial",
                             "log.norm.const", "log.prior", "quit"),
                     theta = NULL, args = NULL)
{
    interpret.theta = function(n, theta)
        return (list(prec = exp(theta[1L])))
    graph = function(n, theta)
        return (Diagonal(n, x= rep(1, n)))
    Q = function(n, theta) {
        prec = interpret.theta(n, theta)$prec
        return (Diagonal(n, x= rep(prec, n))) }
    mu = function(n, theta) return (numeric(0))
    log.norm.const = function(n, theta) {
        prec = interpret.theta(n, theta)$prec
        return (sum(dnorm(rep(0, n),
                          sd = 1/sqrt(prec), log=TRUE))) }
    log.prior = function(n, theta) {
        prec = interpret.theta(n, theta)$prec
        return (dgamma(prec, shape = 1, rate = 5e-05, log=TRUE)
                + theta[1L]) }
    initial = function(n, theta) return (4.0)
    quit = function(n, theta) return (invisible())

    val = do.call(match.arg(cmd),
             args = list(n = as.integer(args$n), theta = theta))
    return (val)
}
n = 50 ## the dimension
my.iid = inla.rgeneric.define(iid.model, n=n)
\end{verbatim}
    Hence, we can replace \texttt{f(idx,model="iid")} with our own
    \texttt{R}-implementation, using \texttt{f(idx, model=my.iid)}.
    For details on the format, see \texttt{inla.doc("rgeneric")} and
    \texttt{demo(rgeneric)}.
\end{description}

\section{A CHALLENGE FOR THE FUTURE: PRIORS}
\label{sec:priors}

Although the \RINLA project has been highly successful, it has also
revealed some ``weak points'' in general Bayesian methodology from a
practical point of view. In particular, our main concern is how we
think about and specify priors in LGMs. We will now discuss this issue
and our current plan to provide good sensible ``default'' priors.

Bayesian statistical models require prior distributions for all the
random elements of the model. Working within the class of LGMs, this
involves choosing \emph{priors} for all the hyperparameters
\mm{\theta} in the model, since the latent field is by definition
Gaussian. We deliberately wrote prior\emph{s} since it is common
practice to define independent priors for each $\theta_j$, while what
we really should aim for is a joint prior for all \mm{\theta}, when
appropriate.

The ability to incorporate prior knowledge in Bayesian statistics is a
great tool and potentially very useful. However, except for cases
where we do have ``real/experimental'' prior knowledge, for example
through results from previous experiments, it is often conceptually
difficult to encode prior knowledge through probability distributions
for all model parameters. Examples include priors for precision and
overdispersion parameters, or the amount of t-ness in the Student-t
distribution. \cite{art631} discuss these aspects in great detail.

In \RINLA we have chosen to provide default prior distributions for
all parameters. We admit that currently these have been chosen partly
based on the priors that are commonly used in the literature and
partly out of the blue. It might be argued that this is not a good
strategy, and that we should force the user to provide the complete
model including the joint prior. This is a valid point, but all priors
in \RINLA can easily be changed, allowing the user to define any
arbitrary prior distribution. So the whole argument boils down to a
question of convenience.

Do we have a ``Houston, we have a problem''-situation with priors?
Looking at the current practice within the Bayesian society, we came
to the conclusion; we do. We will argue for this through a simple
example, showing what can go wrong, how we can think about the problem
and how we can fix it. We only discuss proper priors.

Consider the problem of replacing a linear effect of the Townsend
deprivation index \texttt{tpi} with a smooth effect of \texttt{tpi} in
the Leukaemia example in \Sec{sec:spatial}. This is easily implemented
by replacing \verb|tpi| with \verb|f(tpi, model="rw2")|. Here,
\verb|rw2| is a stochastic spline, simply saying that the second
derivative is independent Gaussian noise \citep{book80,art435}. By
default, we constrain the smooth effect to also sum to zero, so that
these two model formulations are the same in the limit as the
precision parameter $\tau$ tends to infinity, and a vague Gaussian
prior is used for the linear effect. The question is which prior
should be used for $\tau$. An overwhelming majority of cases in the
literature uses some kind of a Gamma$(a,b)$ prior for $\tau$, implying
that $\pi(\tau)\propto \tau^{a-1}\exp(-b\tau)$, for some $a,b>0$. This
prior is flexible, conjugate with the Gaussian, and seems like a
convenient choice. Since almost everyone else is using it, how wrong
can it be?

\begin{figure}
    \centering
    \begin{minipage}[b]{0.49\linewidth}
        \includegraphics[width=\linewidth]{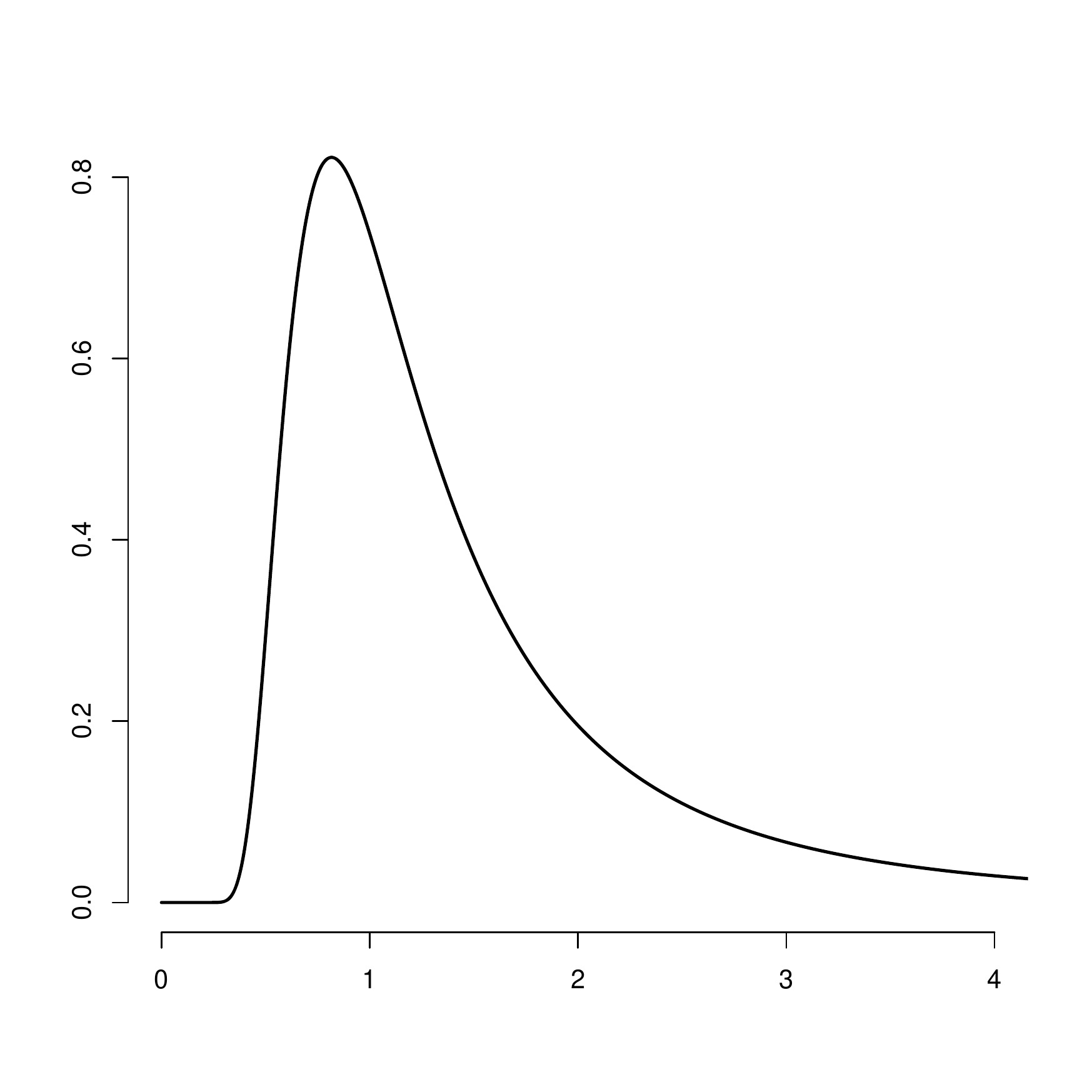}
        \vspace*{-1cm}\center{(a)}
    \end{minipage}
    \begin{minipage}[b]{0.49\linewidth}
        \includegraphics[width=\linewidth]{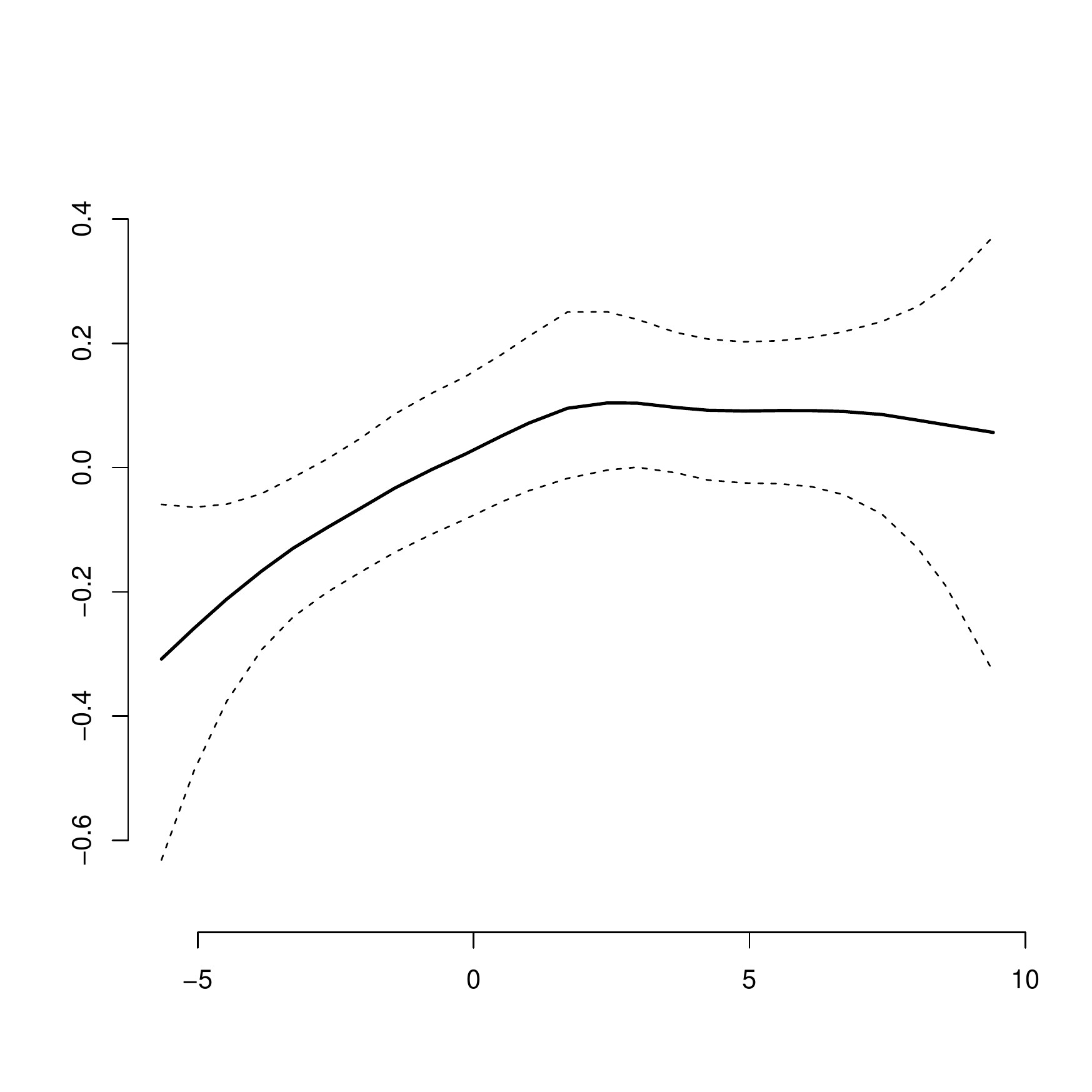}
        \vspace*{-1cm}\center{(b)}
    \end{minipage}
    \vspace*{5mm}
    \caption{Panel (a) shows the Gamma$(1,1)$ prior on the distance
        scale. Panel (b) shows the smoothed effect of covariate
        \texttt{tpi} using the exponential prior on the distance scale
        $\lambda\exp(-\lambda )$.}
    \label{fig7}
\end{figure}

If we rewind to the point where we replaced the linear effect with a
smooth effect, we realise that we do this because we want a more
flexible model than the linear effect, i.e.\ we also want to capture
\emph{deviations} from the linear effect. Implicitly, \emph{if} there
is a linear effect, we do want to retrieve that with enough data.
Measuring the distance between the straight line and the stochastic
spline using the Kullback-Leibler divergence, we find that
$\text{KLD} \propto 1/\tau$ meaning that the (unidirectional) distance
is $d \propto\sqrt{1/\tau}$. For simplicity, choose $a=b=1$ in the
Gamma-prior, then the derived prior for the distance $d$ is
\begin{equation}\label{eq19}%
    \pi(d) \propto \exp(-1/d^{2})/d^{3}.
\end{equation}
\Fig{fig7}a displays this prior on the distance scale, revealing two
surprising features. First, the mode is around $d\approx0.82$, and
second, the prior appears to be zero for a range of positive
distances. The second feature is serious as it simply \emph{prevents}
the spline from getting too close to the linear effect. It is clear
from \eref{eq19} that the effect is severe, and in practice,
$\pi(d) \approx 0$ even for positive $d$. This is an example of what
\cite{art631} call prior \emph{overfitting}; the prior prevents the
simpler model to be located, even when it is the true model.
Choosing different parameters in the Gamma-prior does not change the
overfitting issue. For all $a,b>0$, the corresponding prior for the
distance tends to $0$ as $d\rightarrow 0$. For a (well-behaved) prior
to have $\pi(d=0) > 0$, we need $E(\tau)=\infty$.

If we are concerned about the behaviour of the distance between the
more flexible and the simpler model component, we should define the
prior directly on the distance, as proposed in \cite{art631}. A prior
for the distance should be decaying with the mode at distance zero.
This makes the simpler model central and the point of attraction. The
exponential prior is recommended as a generic choice since it has a
constant rate penalisation, $\pi(d) = \lambda \exp(-\lambda d)$. The
value of $\lambda$ could be chosen by calibrating some property of the
model component under consideration. Note that this way of defining
the prior is invariant to reparameterisations, as it is defined on the
distance and not for a particular parametersation.

Let us return to the stochastic spline example, assigning the
exponential prior to the distance. The parameter $\lambda$ can be
calibrated by imposing the knowledge that the effect of \texttt{tpi}
is not likely to be above $1$ on the linear predictor scale,
\begin{verbatim}
..+ f(tpi, model="rw2", scale.model = TRUE,
      hyper = list(prec = list(prior="pc.prec",  param=c(1, 0.01))))
\end{verbatim}
Here, \texttt{scale.model} is required to ensure that the parameter
$\tau$ represents the precision, not just \emph{a} precision parameter
\citep{art521}. The estimated results are given in \Fig{fig7}b,
illustrating the point-wise posterior mean, median and the $2.5\%$ and
$97.5\%$ credibility intervals, for the effect of \texttt{tpi} on the
mean survival time.

Here, we have only briefly addressed the important topic of
constructing well-working priors, and currently we are focusing a lot
of activity on this issue to take the development further. Besides
others we plan to integrate automatic tests for prior sensitivity,
following the work of \citet{roos-held-2011,art573}. The final goal is
to use the above ideas to construct a joint default prior for LGMs,
which can be easily understood and interpreted. A main issue is how to
decompose and control the variance of the linear predictor, an issue
we have not discussed here. For further information about this issue,
please see \cite{art631} for the original report which introduces the
class of penalised complexity (PC) priors. Some examples on
application of these priors include disease mapping \citep{art585},
bivariate meta-analysis \citep{art586,guo-riebler-2015},
age-period-cohort models \citep{riebler-held-2016}, Bayesian P-splines
\citep{art630}, structured additive distributional regression
\citep{klein-kneib-2016}, Gaussian fields in spatial statistics
\citep{art590}, modeling monthly maxima of instantaneous flow
\citep{ferkingstad-etal-2016} and autoregressive processes
\citep{tech126}.

Interestingly, the framework and ideas of PC priors, are also useful
for sensitivity analysis of model assumptions and developing robust
models, but it is too early to report this here. Stay tuned!

\section{DISCUSSION}
\label{sec:discussion}

We hope we have convinced the reader that the INLA approach to
approximate Bayesian inference for LGMs is a useful addition to the
applied statistician's toolbox; the key components just play so nicely
together, providing a very exact approximation while reducing
computation costs substantially. The key benefit of the INLA approach
is that it is central to our long-term goal of making LGMs a class of
models that we (as a community) can \emph{use and understand}.

Developing, writing and maintaining the code-base for a such large
open-source project, is a huge job. Nearly all the \texttt{R/C/C++}
code is written and maintained by F.\ Lindgren (20\%) and H.\ Rue
(80\%), and is a result of a substantial amount of work over many
years. Many more have contributed indirectly by challenging the
current practice and implementation. The current version of this
project is a result of the cumulative effort of the many users, and
their willingness to share, challenge and question essentially
everything. Documentation is something we could and should improve
upon, but the recent book by \cite{book125} does a really good job.

The current status of the package is good, but we have to account for
the fact that the software has been developed over many years, and is
basically the version we used while developing the methods. Hence,
while the software works well it less streamlined and less easy to
maintain than it ought to be. We are now at a stage where we know what
we want the package to do and software to be, hence a proper rewrite
by skilled people would really be a useful project for the society. If
this would happen, we would be more than happy to share all our
knowledge into a such ``version 2.0'' project!

Another use of \RINLA is to use it purely as computational back-end.
The generality of \RINLA comes with a prize of complexity for the
user, hence a simplified interface for a restricted set of models can
be useful to improve accessibility for a specific target audience or
provide additional tools that are mainly relevant for these models.
Examples of such projects, are \texttt{AnimalINLA} \citep{art622},
\texttt{ShrinkBayes} \citep{art624,art514,art623,riebler-etal-2014},
\texttt{meta4diag} \citep{guo-riebler-2015}, \texttt{BAPC}
\citep{riebler-held-2016}, \texttt{diseasemapping} and
\texttt{geostatp} \citep{art625}, and \cite{art528}. Similarly, the
\texttt{excursions} package for calculating joint exceedance
probabilities in GMRFs \citep{bolin-lindgren-2015,bolin-lindgren-2016}
includes an interface to analyse LGMs estimated by \texttt{R-INLA}.
Recent work on
methodology for filtered spatial point patterns in the context of
distance sampling \citep{art589} has initiated the construction of
wrapper software for fitting other complex spatial models such as
those resulting from plot sampling data or for point process models
within \RINLA. There is also an interesting line of research using
\RINLA to do approximate inference on a sub-model within a larger
model, see \cite{art407} for a theoretical justification and
\cite{art503} for an early application of this idea. One particular
application here, is how to handle missing data in cases where the
joint model is not an LGM.

Please visit us at \texttt{www.r-inla.org}!

\section*{ACKNOWLEDGEMENTS}

We would like to acknowledge all the users of the \RINLA package,
who have challenged and questioned essentially everything, and their
willingness to share this with us.

\small\bibliography{mybib,local-bib}

\begin{thebibliography}{}

\bibitem[Azzalini and Capitanio, 1999]{art414}
Azzalini, A. and Capitanio, A. (1999).
\newblock Statistical applications of the multivariate skew-normal
  distribution.
\newblock {\em Journal of the Royal Statistical Society, Series B},
  61(4):579--602.

\bibitem[Baghishani and Mohammadzadeh, 2012]{art619}
Baghishani, H. and Mohammadzadeh, M. (2012).
\newblock Asymptotic normality of posterior distributions for generalized
  linear mixed models.
\newblock {\em Journal of Multivariate Analysis}, 111:66 -- 77.

\bibitem[Bakka et~al., 2016]{tech125}
Bakka, H., Vanhatalo, J., Illian, J., Simpson, D., and Rue, H. (2016).
\newblock Accounting for physical barriers in species distribution modeling
  with non-stationary spatial random effects.
\newblock arXiv preprint arXiv:1608.03787, {N}orwegian University of Science
  and Technology, {T}rondheim, {N}orway.

\bibitem[{Barndorff-Nielsen} and Cox, 1989]{book123}
{Barndorff-Nielsen}, O.~E. and Cox, D.~R. (1989).
\newblock {\em Asymptotic Techniques for Use in Statistics}, volume~31 of {\em
  Monographs on Statistics and Applied Probability}.
\newblock Chapman and Hall/CRC.

\bibitem[Bauer et~al., 2016]{art582}
Bauer, C., Wakefield, J., Rue, H., Self, S., Feng, Z., and Wang, Y. (2016).
\newblock Bayesian penalized spline models for the analysis of spatio-temporal
  count data.
\newblock {\em Statistics in Medicine}, 35(11):1848--1865.

\bibitem[Bhatt et~al., 2015]{art618}
Bhatt, S., Weiss, D.~J., Cameron, E., Bisanzio, D., Mappin, B., Dalrymple, U.,
  Battle, K.~E., Moyes, C.~L., Henry, A., Eckhoff, P.~A., Wenger, E.~A.,
  Briët, O., Penny, M.~A., Smith, T.~A., Bennett, A., Yukich, J., Eisele,
  T.~P., Griffin, J.~T., Fergus, C.~A., Lynch, M., Lindgren, F., Cohen, J.~M.,
  Murray, C. L.~J., Smith, D.~L., Hay, S.~I., Cibulskis, R.~E., and Gething,
  P.~W. (2015).
\newblock The effect of malaria control on plasmodium falciparum in {A}frica
  between 2000 and 2015.
\newblock {\em Nature}, (526):207--211.

\bibitem[Bivand et~al., 2015]{art528}
Bivand, R.~S., {G\'omez-Rubio}, V., and Rue, H. (2015).
\newblock Spatial data analysis with {R-INLA} with some extensions.
\newblock {\em Journal of Statistical Software}, 63(20):1--31.

\bibitem[Blangiardo and Cameletti, 2015]{book125}
Blangiardo, M. and Cameletti, M. (2015).
\newblock {\em Spatial and Spatio-temporal {B}ayesian Models with {R-INLA}}.
\newblock John Wiley \& Sons.

\bibitem[Bolin and Lindgren, 2015]{bolin-lindgren-2015}
Bolin, D. and Lindgren, F. (2015).
\newblock Excursion and contour uncertainty regions for latent {G}aussian
  models.
\newblock {\em Journal of the Royal Statistical Society, Series B},
  77(1):85--106.

\bibitem[Bolin and Lindgren, 2016]{bolin-lindgren-2016}
Bolin, D. and Lindgren, F. (2016).
\newblock Quantifying the uncertainty of contour maps.
\newblock {\em Journal of Computational and Graphical Statistics}.
\newblock arXiv preprint arXiv:1507.01778, to appear.

\bibitem[Bowler et~al., 2015]{art611}
Bowler, D.~E., Haase, P., Kr{\"o}ncke, I., Tackenberg, O., Bauer, H.~G.,
  Brendel, C., Brooker, R.~W., Gerisch, M., Henle, K., Hickler, T., Hof, C.,
  Klotz, S., K{\"u}hn, I., Matesanz, S., {O‘Hara}, R., Russell, D.,
  Schweiger, O., Valladares, F., Welk, E., Wiemers, M., and
  {B{\"o}hning-Gaese}, K. (2015).
\newblock A cross-taxon analysis of the impact of climate change on abundance
  trends in central {E}urope.
\newblock {\em Biological Conservation}, 187:41--50.

\bibitem[Box and Tiao, 1973]{book77}
Box, G. E.~P. and Tiao, G.~C. (1973).
\newblock {\em {B}ayesian Inference in Statistical Analysis}.
\newblock Addison-Wesley Publishing Co., Reading, Mass.-London-Don Mills, Ont.

\bibitem[Box and Wilson, 1951]{art400}
Box, G. E.~P. and Wilson, K.~B. (1951).
\newblock On the experimental attainment of optimum conditions (with
  discussion).
\newblock {\em Journal of the Royal Statistical Society, Series B},
  13(1):1--45.

\bibitem[Brown, 2015]{art625}
Brown, P.~E. (2015).
\newblock Model-based geostatistics the easy way.
\newblock {\em Journal of Statistical Software}, 63(12):1--24.

\bibitem[Crewe and Mccracken, 2015]{art606}
Crewe, T.~L. and Mccracken, J.~D. (2015).
\newblock Long-term trends in the number of monarch butterflies ({L}epidoptera:
  {N}ymphalidae) counted on fall migration at {L}ong {P}oint, {O}ntario,
  {C}anada (1995{\textendash}2014).
\newblock {\em Annals of the Entomological Society of America}.

\bibitem[{Dwyer-Lindgren} et~al., 2015]{art612}
{Dwyer-Lindgren}, L., Flaxman, A.~D., Ng, M., Hansen, G.~M., Murray, C.~J., and
  Mokdad, A.~H. (2015).
\newblock Drinking patterns in {US} counties from 2002 to 2012.
\newblock {\em American Journal of Public Health}, 105(6):1120--1127.

\bibitem[Ferkingstad et~al., 2016]{ferkingstad-etal-2016}
Ferkingstad, E., Geirsson, O.~P., Hrafnkelsson, B., Davidsson, O.~B., and
  Gardarsson, S.~M. (2016).
\newblock A {B}ayesian hierarchical model for monthly maxima of instantaneous
  flow.
\newblock {\em arXiv preprint arXiv:1606.07667}.

\bibitem[Ferkingstad and Rue, 2015]{art587}
Ferkingstad, E. and Rue, H. (2015).
\newblock Improving the {INLA} approach for approximate {B}ayesian inference
  for latent {G}aussian models.
\newblock {\em Electronic Journal of Statistics}, 9:2706--2731.

\bibitem[Friedrich et~al., 2016]{art603}
Friedrich, A., Marshall, J.~C., Biggs, P.~J., Midwinter, A.~C., and French,
  N.~P. (2016).
\newblock Seasonality of campylobacter jejuni isolates associated with human
  campylobacteriosis in the {M}anawatu region, {N}ew {Z}ealand.
\newblock {\em Epidemiology and Infection}, 144:820--828.

\bibitem[Fuglstad et~al., 2015a]{art529}
Fuglstad, G.~A., Lindgren, F., Simpson, D., and Rue, H. (2015a).
\newblock Exploring a new class of non-stationary spatial {G}aussian random
  fields with varying local anisotropy.
\newblock {\em Statistica Sinica}, 25(1):115--133.
\newblock Special issue of {Spatial} and {Temporal} {Data} {Analysis}.

\bibitem[Fuglstad et~al., 2015b]{art581}
Fuglstad, G.~A., Simpson, D., Lindgren, F., and Rue, H. (2015b).
\newblock Does non-stationary spatial data always require non-stationary random
  fields?
\newblock {\em Spatial Statistics}, 14, Part C:505--531.

\bibitem[Fuglstad et~al., 2016]{art590}
Fuglstad, G.~A., Simpson, D., Lindgren, F., and Rue, H. (2016).
\newblock Constructing priors that penalize the complexity of {G}aussian random
  fields.
\newblock {\em Submitted}, xx(xx):xx--xx.
\newblock arXiv:1503.00256.

\bibitem[{Garc{\'i}a-P{\'e}rez} et~al., 2015]{art609}
{Garc{\'i}a-P{\'e}rez}, J., Lope, V., {L{\'o}pez-Abente}, G.,
  {Gonz{\'a}lez-S{\'a}nchez}, M., and {Fern{\'a}ndez-Navarro}, P. (2015).
\newblock Ovarian cancer mortality and industrial pollution.
\newblock {\em Environmental Pollution}, 205:103 -- 110.

\bibitem[Gelman et~al., 2014]{art627}
Gelman, A., Hwang, J., and Vehtari, A. (2014).
\newblock Understanding predictive information criteria for {B}ayesian models.
\newblock {\em Statistics and Computing}, 24(6):997--1016.

\bibitem[Gneiting and Raftery, 2007]{art427}
Gneiting, T. and Raftery, A.~E. (2007).
\newblock Strictly proper scoring rules, prediction, and estimation.
\newblock {\em Journal of the American Statistical Association}, 102:359--378.

\bibitem[Goicoa et~al., 2016]{art592}
Goicoa, T., Ugarte, M.~D., Etxeberria, J., and Militino, A.~F. (2016).
\newblock Age-space-time {CAR} models in {B}ayesian disease mapping.
\newblock {\em Statistics in Medicine}, 35(14):2391--2405.
\newblock sim.6873.

\bibitem[Goth et~al., 2014]{art600}
Goth, U.~S., Hammer, H.~L., and Claussen, B. (2014).
\newblock Utilization of {N}orway’s emergency wards: The second 5 years after
  the introduction of the patient list system.
\newblock {\em International Journal of Environmental Research and Public
  Health}, 11(3):3375.

\bibitem[Guihenneuc-{J}ouyaux and Rousseau, 2005]{art407}
Guihenneuc-{J}ouyaux, C. and Rousseau, J. (2005).
\newblock Laplace expansion in {M}arkov chain {M}onte {C}arlo algorithms.
\newblock {\em Journal of Computational and Graphical Statistics},
  14(1):75--94.

\bibitem[Guo and Riebler, 2015]{guo-riebler-2015}
Guo, J. and Riebler, A. (2015).
\newblock meta4diag: Bayesian bivariate meta-analysis of diagnostic test
  studies for routine practice.
\newblock {\em arXiv preprint arXiv:1512.06220}.

\bibitem[Guo et~al., 2015]{art586}
Guo, J., Riebler, A., and Rue, H. (2015).
\newblock Bayesian bivariate meta-analysis of diagnostic test studies with
  interpretable priors.
\newblock {\em In revision for Statistics in Medicine}, xx(xx):xx--xx.
\newblock arXiv:1512.06217.

\bibitem[Halonen et~al., 2016]{art597}
Halonen, J.~I., Blangiardo, M., Toledano, M.~B., Fecht, D., Gulliver, J.,
  Anderson, H.~R., Beevers, S.~D., Dajnak, D., Kelly, F.~J., and Tonne, C.
  (2016).
\newblock Long-term exposure to traffic pollution and hospital admissions in
  {L}ondon.
\newblock {\em Environmental Pollution}, 208, Part A:48 -- 57.
\newblock Special Issue: Urban Health and Wellbeing.

\bibitem[Halonen et~al., 2015]{art607}
Halonen, J.~I., Hansell, A.~L., Gulliver, J., Morley, D., Blangiardo, M.,
  Fecht, D., Toledano, M.~B., Beevers, S.~D., Anderson, H.~R., Kelly, F.~J.,
  and Tonne, C. (2015).
\newblock Road traffic noise is associated with increased cardiovascular
  morbidity and mortality and all-cause mortality in {L}ondon.
\newblock {\em European Heart Journal}.

\bibitem[Held and Rue, 2010]{col26}
Held, L. and Rue, H. (2010).
\newblock Conditional and intrinsic autoregressions.
\newblock In Gelfand, A., Diggle, P., Fuentes, M., and Guttorp, P., editors,
  {\em Handbook of Spatial Statistics}, pages 201--216. CRC/Chapman \& Hall,
  Boca Raton, FL.

\bibitem[Held and Sauter, 2016]{held-sauter-2016}
Held, L. and Sauter, R. (2016).
\newblock Adaptive prior weighting in generalized regression.
\newblock {\em Biometrics}.
\newblock To appear.

\bibitem[Held et~al., 2010]{col28}
Held, L., Schr{\"o}dle, B., and Rue, H. (2010).
\newblock Posterior and cross-validatory predictive checks: {A} comparison of
  {MCMC} and {INLA}.
\newblock In Kneib, T. and Tutz, G., editors, {\em Statistical Modelling and
  Regression Structures -- Festschrift in Honour of Ludwig Fahrmeir}, pages
  91--110. Springer Verlag, Berlin.

\bibitem[Henderson et~al., 2002]{art458}
Henderson, R., Shimakura, S., and Gorst, D. (2002).
\newblock Modeling spatial variation in leukemia survival data.
\newblock {\em Journal of the American Statistical Association},
  97(460):965--972.

\bibitem[Hodges, 2013]{book124}
Hodges, J.~S. (2013).
\newblock {\em Richly Parameterized Linear Models: Additive, Time Series, and
  Spatial Models Using Random Effects}.
\newblock Chapman \& Hall/CRC Texts in Statistical Science. Chapman and
  Hall/CRC.

\bibitem[Holand et~al., 2013]{art622}
Holand, A.~M., Steinsland, I., Martino, S., and Jensen, H. (2013).
\newblock Animal models and integrated nested {L}aplace approximations.
\newblock {\em {G3: Genes|Genomics|Genetics}}, 3(8):1241--1251.

\bibitem[Hu and Steinsland, 2016]{art591}
Hu, X. and Steinsland, I. (2016).
\newblock Spatial modeling with system of stochastic partial differential
  equations.
\newblock {\em Wiley Interdisciplinary Reviews: Computational Statistics},
  8(2):112--125.

\bibitem[Iulian et~al., 2015]{art608}
Iulian, T.~V., Juan, P., and Mateu, J. (2015).
\newblock Bayesian spatio-temporal prediction of cancer dynamics.
\newblock {\em Computers \& Mathematics with Applications}, 70(5):857--868.

\bibitem[Jousimo et~al., 2014]{art616}
Jousimo, J., Tack, A. J.~M., Ovaskainen, O., Mononen, T., Susi, H., Tollenaere,
  C., and Laine, A.-L. (2014).
\newblock Ecological and evolutionary effects of fragmentation on infectious
  disease dynamics.
\newblock {\em Science}, 344(6189):1289--1293.

\bibitem[Kandt et~al., 2016]{art594}
Kandt, J., Chang, S., Yip, P., and Burdett, R. (2016).
\newblock The spatial pattern of premature mortality in {H}ong {K}ong: How does
  it relate to public housing?
\newblock {\em Urban Studies}.

\bibitem[Karagiannis-Voules et~al., 2015]{art617}
Karagiannis-Voules, D.-A., Biedermann, P., Ekpo, U.~F., Garba, A., Langer, E.,
  Mathieu, E., Midzi, N., Mwinzi, P., Polderman, A.~M., Raso, G., Sacko, M.,
  Talla, I., Tchuent{\'e}, L.-A.~T., Tour{\'e}, S., Winkler, M.~S., Utzinger,
  J., and Vounatsou, P. (2015).
\newblock Spatial and temporal distribution of soil-transmitted helminth
  infection in sub-{S}aharan {A}frica: a systematic review and geostatistical
  meta-analysis.
\newblock {\em The Lancet Infectious Diseases}, 15(1):74 -- 84.

\bibitem[Karcher et~al., 2016]{art610}
Karcher, M.~D., Palacios, J.~A., Bedford, T., Suchard, M.~A., and Minin, V.~N.
  (2016).
\newblock Quantifying and mitigating the effect of preferential sampling on
  phylodynamic inference.
\newblock {\em PLoS Comput Biol}, 12(3):1--19.

\bibitem[Kauermann et~al., 2009]{art620}
Kauermann, G., Krivobokova, T., and Fahrmeir, L. (2009).
\newblock Some asymptotic results on generalized penalized spline smoothing.
\newblock {\em Journal of the Royal Statistical Society: Series B (Statistical
  Methodology)}, 71(2):487--503.

\bibitem[Klein and Kneib, 2016]{klein-kneib-2016}
Klein, N. and Kneib, T. (2016).
\newblock Scale-dependent priors for variance parameters in structured additive
  distributional regression.
\newblock {\em Bayesian Analysis}.
\newblock To appear.

\bibitem[Kr{\"o}ger et~al., 2016]{art601}
Kr{\"o}ger, H., Hoffmann, R., and Pakpahan, E. (2016).
\newblock Consequences of measurement error for inference in cross-lagged panel
  design-the example of the reciprocal causal relationship between subjective
  health and socio-economic status.
\newblock {\em Journal of the Royal Statistical Society, Series A},
  179(2):607--628.

\bibitem[Li et~al., 2012]{art503}
Li, Y., Brown, P., Rue, H., {al-Maini}, M., and Fortin, P. (2012).
\newblock Spatial modelling of {Lupus} incidence over 40 years with changes in
  census areas.
\newblock {\em Journal of the Royal Statistical Society, Series C}, 61:99--115.

\bibitem[Lindgren and Rue, 2008]{art435}
Lindgren, F. and Rue, H. (2008).
\newblock A note on the second order random walk model for irregular locations.
\newblock {\em Scandinavian Journal of Statistics}, 35(4):691--700.

\bibitem[Lindgren and Rue, 2015]{art527}
Lindgren, F. and Rue, H. (2015).
\newblock Bayesian spatial modelling with {R-INLA}.
\newblock {\em Journal of Statistical Software}, 63(19):1--25.

\bibitem[Lindgren et~al., 2011]{art500}
Lindgren, F., Rue, H., and Lindstr{\"o}m, J. (2011).
\newblock An explicit link between {G}aussian fields and {G}aussian {M}arkov
  random fields: {T}he {SPDE} approach (with discussion).
\newblock {\em Journal of the Royal Statistical Society, Series B},
  73(4):423--498.

\bibitem[Lithio and Nettleton, 2015]{art599}
Lithio, A. and Nettleton, D. (2015).
\newblock Hierarchical modeling and differential expression analysis for
  {RNA}-seq experiments with inbred and hybrid genotypes.
\newblock {\em Journal of Agricultural, Biological, and Environmental
  Statistics}, 20(4):598--613.

\bibitem[Martino et~al., 2010]{art494}
Martino, S., Akerkar, R., and Rue, H. (2010).
\newblock Approximate {B}ayesian inference for survival models.
\newblock {\em Scandinavian Journal of Statistics}, 28(3):514--528.

\bibitem[Martins and Rue, 2014]{art531}
Martins, T.~G. and Rue, H. (2014).
\newblock Extending {INLA} to a class of near-{G}aussian latent models.
\newblock {\em Scandinavian Journal of Statistics}, 41(4):893--912.

\bibitem[Martins et~al., 2013]{art522}
Martins, T.~G., Simpson, D., Lindgren, F., and Rue, H. (2013).
\newblock Bayesian computing with {INLA}: {N}ew features.
\newblock {\em Computational Statistics \& Data Analysis}, 67:68--83.

\bibitem[Muff and Keller, 2015]{muff-keller-2015}
Muff, S. and Keller, L.~F. (2015).
\newblock Reverse attenuation in interaction terms due to covariate measurement
  error.
\newblock {\em Biometrical Journal}, 57(6):1068--1083.

\bibitem[Muff et~al., 2015]{art561}
Muff, S., Riebler, A., Rue, H., Saner, P., and Held, L. (2015).
\newblock Bayesian analysis of measurement error models using integrated nested
  {L}aplace approximations.
\newblock {\em Journal of the Royal Statistical Society, Series C},
  64(2):231--252.

\bibitem[Niemi et~al., 2015]{art596}
Niemi, J., Mittman, E., Landau, W., and Nettleton, D. (2015).
\newblock Empirical {B}ayes analysis of {RNA}-seq data for detection of gene
  expression heterosis.
\newblock {\em Journal of Agricultural, Biological, and Environmental
  Statistics}, 20(4):614--628.

\bibitem[Noor et~al., 2014]{art615}
Noor, A.~M., Kinyoki, D.~K., Mundia, C.~W., Kabaria, C.~W., Mutua, J.~W.,
  Alegana, V.~A., Fall, I.~S., and Snow, R.~W. (2014).
\newblock The changing risk of {P}lasmodium falciparum malaria infection in
  {A}frica: 2000-10: a spatial and temporal analysis of transmission intensity.
\newblock {\em The Lancet}, 383(9930):1739--1747.

\bibitem[{Ogden}, 2016]{art621}
{Ogden}, H. (2016).
\newblock {On asymptotic validity of approximate likelihood inference}.
\newblock {\em ArXiv e-prints}.

\bibitem[Opitz et~al., 2016]{art598}
Opitz, N., Marcon, C., Paschold, A., Malik, W.~A., Lithio, A., Brandt, R.,
  Piepho, H., Nettleton, D., and Hochholdinger, F. (2016).
\newblock Extensive tissue-specific transcriptomic plasticity in maize primary
  roots upon water deficit.
\newblock {\em Journal of Experimental Botany}, 67(4):1095--1107.

\bibitem[Papoila et~al., 2014]{papoila-etal-2014}
Papoila, A.~L., Riebler, A., Amaral-Turkman, A., S{\~a}o-Jo{\~a}o, R., Ribeiro,
  C., Geraldes, C., and Miranda, A. (2014).
\newblock Stomach cancer incidence in {S}outhern {P}ortugal 1998--2006: A
  spatio-temporal analysis.
\newblock {\em Biometrical Journal}, 56(3):403--415.

\bibitem[Plummer, 2016]{man2}
Plummer, M. (2016).
\newblock {\em rjags: {B}ayesian Graphical Models using {MCMC}}.
\newblock R package version 4-6.

\bibitem[Quiroz et~al., 2015]{art549}
Quiroz, Z., Prates, M.~O., and Rue, H. (2015).
\newblock A {B}ayesian approach to estimate the biomass of anchovies in the
  coast of {Per\'u}.
\newblock {\em Biometrics}, 71(1):208--217.

\bibitem[Riebler and Held, 2016]{riebler-held-2016}
Riebler, A. and Held, L. (2016).
\newblock Projecting the future burden of cancer: {B}ayesian age-period-cohort
  analysis with integrated nested {L}aplace approximations.
\newblock {\em Biometrical Journal}.
\newblock Conditionally accepted.

\bibitem[Riebler et~al., 2012]{art492}
Riebler, A., Held, L., and Rue, H. (2012).
\newblock Estimation and extrapolation of time trends in registry data -
  {B}orrowing strength from related populations.
\newblock {\em Annals of Applied Statistics}, 6(1):304--333.

\bibitem[Riebler et~al., 2014]{riebler-etal-2014}
Riebler, A., Robinson, M., and van~de Wiel, M. (2014).
\newblock {\em Statistical Analysis of Next Generation Sequencing Data},
  chapter Analysis of Next Generation Sequencing Data Using Integrated Nested
  Laplace Approximation (INLA), pages 75--91.
\newblock Springer International Publishing.

\bibitem[Riebler et~al., 2016]{art585}
Riebler, A., S{\o}rbye, S.~H., Simpson, D., and Rue, H. (2016).
\newblock An intuitive {B}ayesian spatial model for disease mapping that
  accounts for scaling.
\newblock {\em Statistical Methods in Medical Research}, 25(4):1145--1165.

\bibitem[Robert and Casella, 1999]{book44}
Robert, C.~P. and Casella, G. (1999).
\newblock {\em {M}onte {C}arlo Statistical Methods}.
\newblock Springer-Verlag, New York.

\bibitem[Rooney et~al., 2015]{art614}
Rooney, J., Vajda, A., Heverin, M., Elamin, M., Crampsie, A., McLaughlin, R.,
  Staines, A., and Hardiman, O. (2015).
\newblock Spatial cluster analysis of population amyotrophic lateral sclerosis
  risk in {Ireland}.
\newblock {\em Neurology}, 84:1537--1544.

\bibitem[Roos and Held, 2011]{roos-held-2011}
Roos, M. and Held, L. (2011).
\newblock Sensitivity analysis in {B}ayesian generalized linear mixed models
  for binary data.
\newblock {\em Bayesian Analysis}, 6(2):259--278.

\bibitem[Roos et~al., 2015a]{art573}
Roos, M., Martins, T.~G., Held, L., and Rue, H. (2015a).
\newblock Sensitivity analysis for {B}ayesian hierarchical models.
\newblock {\em Bayesian Analysis}, 10(2):321--349.

\bibitem[Roos et~al., 2015b]{art604}
Roos, N.~C., Carvalho, A.~R., Lopes, P.~F., and Pennino, M.~G. (2015b).
\newblock Modeling sensitive parrotfish ({L}abridae: {S}carini) habitats along
  the {B}razilian coast.
\newblock {\em Marine Environmental Research}, 110:92 -- 100.

\bibitem[Rue and Held, 2005]{book80}
Rue, H. and Held, L. (2005).
\newblock {\em Gaussian {M}arkov Random Fields: {T}heory and Applications},
  volume 104 of {\em Monographs on Statistics and Applied Probability}.
\newblock Chapman \& Hall, London.

\bibitem[Rue and Held, 2010]{col27}
Rue, H. and Held, L. (2010).
\newblock Markov random fields.
\newblock In Gelfand, A., Diggle, P., Fuentes, M., and Guttorp, P., editors,
  {\em Handbook of Spatial Statistics}, pages 171--200. CRC/Chapman \& Hall,
  Boca Raton, FL.

\bibitem[Rue et~al., 2009]{art451}
Rue, H., Martino, S., and Chopin, N. (2009).
\newblock Approximate {B}ayesian inference for latent {G}aussian models using
  integrated nested {L}aplace approximations (with discussion).
\newblock {\em Journal of the Royal Statistical Society, Series B},
  71(2):319--392.

\bibitem[Salmon et~al., 2015]{art605}
Salmon, M., Schumacher, D., Stark, K., and H{\"o}hle, M. (2015).
\newblock Bayesian outbreak detection in the presence of reporting delays.
\newblock {\em Biometrical Journal}, 57(6):1051--1067.

\bibitem[Santermans et~al., 2016]{art593}
Santermans, E., Robesyn, E., Ganyani, T., Sudre, B., Faes, C., Quinten, C.,
  Van~Bortel, W., Haber, T., Kovac, T., Van~Reeth, F., Testa, M., Hens, N., and
  Plachouras, D. (2016).
\newblock Spatiotemporal evolution of ebola virus disease at sub-national level
  during the 2014 {W}est {A}frica epidemic: Model scrutiny and data meagreness.
\newblock {\em PLoS ONE}, 11(1):1--11.

\bibitem[Sauter and Held, 2015]{art602}
Sauter, R. and Held, L. (2015).
\newblock Network meta-analysis with integrated nested {L}aplace
  approximations.
\newblock {\em Biometrical Journal}, 57(6):1038--1050.

\bibitem[Sauter and Held, 2016]{sauter-held-2016}
Sauter, R. and Held, L. (2016).
\newblock Quasi-complete separation in random effects of binary response mixed
  models.
\newblock {\em Journal of Statistical Computation and Simulation},
  86(14):2781--2796.

\bibitem[Schr{\"o}dle and Held, 2011a]{schroedle-held-2011b}
Schr{\"o}dle, B. and Held, L. (2011a).
\newblock A primer on disease mapping and ecological regression using {INLA}.
\newblock {\em Computational Statistics}, 26(2):241--258.

\bibitem[Schr{\"o}dle and Held, 2011b]{schroedle-held-2011}
Schr{\"o}dle, B. and Held, L. (2011b).
\newblock Spatio-temporal disease mapping using {INLA}.
\newblock {\em Environmetrics}, 22(6):725--734.

\bibitem[Schr\"odle et~al., 2012]{art498}
Schr\"odle, B., Held, L., and Rue, H. (2012).
\newblock Assessing the impact of network data on the spatio-temporal spread of
  infectious diseases.
\newblock {\em Biometrics}, 68(3):736--744.

\bibitem[Selwood et~al., 2015]{art613}
Selwood, K.~E., Thomson, J.~R., Clarke, R.~H., McGeoch, M.~A., and Mac~Nally,
  R. (2015).
\newblock Resistance and resilience of terrestrial birds in drying climates: do
  floodplains provide drought refugia?
\newblock {\em Global Ecology and Biogeography}, 24(7):838--848.

\bibitem[Shun and {McC}ullagh, 1995]{art408}
Shun, Z. and {McC}ullagh, P. (1995).
\newblock Laplace approximation of high dimensional integrals.
\newblock {\em Journal of the Royal Statistical Society, Series B},
  57(4):749--760.

\bibitem[Simpson et~al., 2016a]{art583}
Simpson, D., Illian, J., Lindgren, F., S{\o}rbye, S., and Rue, H. (2016a).
\newblock Going off grid: {C}omputational efficient inference for
  log-{G}aussian {C}ox processes.
\newblock {\em Biometrika}, 103(1):1--22.
\newblock (doi: 10.1093/biomet/asv064).

\bibitem[Simpson et~al., 2012]{art508}
Simpson, D., Lindgren, F., and Rue, H. (2012).
\newblock In order to make spatial statistics computationally feasible, we need
  to forget about the covariance function.
\newblock {\em Environmetrics}, 23(1):65--74.

\bibitem[Simpson et~al., 2011]{art512}
Simpson, D.~P., Lindgren, F.~K., and Rue, H. (2011).
\newblock Think continous: {M}arkovian {G}aussian models in spatial statistics.
\newblock {\em Spatial Statistics}, 1(1):16--29.

\bibitem[Simpson et~al., 2016b]{art631}
Simpson, D.~P., Rue, H., Riebler, A., Martins, T.~G., and S{\o}rbye, S.~H.
  (2016b).
\newblock Penalising model component complexity: A principled, practical
  approach to constructing priors (with discussion).
\newblock {\em Statistical Science}, xx(xx):xx--xx.
\newblock (to appear).

\bibitem[S{\o}rbye and Rue, 2014]{art521}
S{\o}rbye, S.~H. and Rue, H. (2014).
\newblock Scaling intrinsic {G}aussian {M}arkov random field priors in spatial
  modelling.
\newblock {\em Spatial Statistics}, 8(3):39--51.

\bibitem[S{\o}rbye and Rue, 2016]{tech126}
S{\o}rbye, S.~H. and Rue, H. (2016).
\newblock Penalised complexity priors for stationary autoregressive processes.
\newblock arXiv preprint arXiv:1608.08941, UiT The {A}rtic {U}niversity of
  {N}orway.

\bibitem[Spiegelhalter et~al., 2002]{art413}
Spiegelhalter, D.~J., Best, N.~G., Carlin, B.~P., and van~der Linde, A. (2002).
\newblock Bayesian measures of model complexity and fit (with discussion).
\newblock {\em Journal of the Royal Statistical Society, Series B},
  64(2):583--639.

\bibitem[Spiegelhalter et~al., 1995]{tech23}
Spiegelhalter, D.~J., Thomas, A., Best, N.~G., and Gilks, W.~R. (1995).
\newblock {BUGS}: {B}ayesian inference using {G}ibbs sampling.
\newblock Version 0.50, MRC {B}iostatistics Unit, Cambridge.

\bibitem[{Stan Development Team}, 2015]{man3}
{Stan Development Team} (2015).
\newblock {\em Stan Modeling Language User's Guide and Reference Manual}.
\newblock Version 2.6.1.

\bibitem[Tierney and Kadane, 1986]{art367}
Tierney, L. and Kadane, J.~B. (1986).
\newblock Accurate approximations for posterior moments and marginal densities.
\newblock {\em Journal of the American Statistical Association},
  81(393):82--86.

\bibitem[Tsiko, 2015]{art595}
Tsiko, R.~G. (2015).
\newblock A spatial latent {G}aussian model for intimate partner violence
  against men in {A}frica.
\newblock {\em Journal of Family Violence}, pages 1--17.

\bibitem[Ugarte et~al., 2016]{ugarte-etal-2016}
Ugarte, M.~D., Adin, A., and Goicoa, T. (2016).
\newblock Two-level spatially structured models in spatio-temporal disease
  mapping.
\newblock {\em Statistical Methods in Medical Research}, 25(4):1080--1100.

\bibitem[Ugarte et~al., 2014]{ugarte-etal-2014}
Ugarte, M.~D., Adin, A., Goicoa, T., and Militino, A.~F. (2014).
\newblock On fitting spatio-temporal disease mapping models using approximate
  {B}ayesian inference.
\newblock {\em Statistical Methods in Medical Research}, 23(6):507--530.

\bibitem[{Van De Wiel} et~al., 2013a]{art624}
{Van De Wiel}, M.~A., {De Menezes}, R.~X., Siebring, E., and {Van Beusechem},
  V.~W. (2013a).
\newblock Analysis of small-sample clinical genomics studies using
  multi-parameter shrinkage: application to high-throughput {RNA} interference
  screening.
\newblock {\em BMC Medical Genomics}, 6(2):1--9.

\bibitem[{Van De Wiel} et~al., 2013b]{art514}
{Van De Wiel}, M.~A., Leday, G. G.~R., Pardo, L., Rue, H., {van der Vaart},
  A.~W., and {van Wieringen}, W.~N. (2013b).
\newblock Bayesian analysis of high-dimensional {RNA} sequencing data:
  estimating priors for shrinkage and multiplicity correction.
\newblock {\em Biostatistics}, 14(1):113--128.

\bibitem[{Van De Wiel} et~al., 2014]{art623}
{Van De Wiel}, M.~A., Neerincx, M., Buffart, T.~E., Sie, D., and Verheul, H.
  M.~W. (2014).
\newblock Shrink{B}ayes: a versatile {R}-package for analysis of count-based
  sequencing data in complex study design.
\newblock {\em BMC Bioinformatics}, 15(1):116.

\bibitem[Ventrucci and Rue, 2016]{art630}
Ventrucci, M. and Rue, H. (2016).
\newblock Penalized complexity priors for degrees of freedom in {B}ayesian
  {P}-splines.
\newblock {\em Statistical Modelling}, xx(xx):xx--xx.
\newblock (to appear).

\bibitem[Wantanabe, 2010]{art626}
Wantanabe, S. (2010).
\newblock Asymptotic equivalence of {B}ayes cross validation and widely
  applicable information criterion in singular learning theory.
\newblock {\em Journal of Machine Learning Research}, 11:3571--3594.

\bibitem[Whittle, 1954]{art246}
Whittle, P. (1954).
\newblock On stationary processes in the plane.
\newblock {\em Biometrika}, 41(3/4):434--449.

\bibitem[Whittle, 1963]{art455}
Whittle, P. (1963).
\newblock Stochastic processes in several dimensions.
\newblock {\em Bull. Inst. Internat. Statist.}, 40:974--994.

\bibitem[Yuan et~al., 2016]{art589}
Yuan, Y., Bachl, F.~E., Borchers, D.~L., Lindgren, F., Illian, J.~B., Buckland,
  S.~T., Rue, H., and Gerrodette, T. (2016).
\newblock Point process models for spatio-temporal distance sampling data.
\newblock {\em SUBMITTED}, xx(xx):xx--xx.
\newblock (arxiv 1604.06013).

\bibitem[Yue et~al., 2014]{art532}
Yue, Y.~R., Simpson, D., Lindgren, F., and Rue, H. (2014).
\newblock Bayesian adaptive smoothing spline using stochastic differential
  equations.
\newblock {\em Bayesian Analysis}, 9(2):397--424.

\end{thebibliography}

\end{document}